\documentclass[a4paper,12pt]{article}

\usepackage[normalem]{ulem}

\newcommand\wordcount{
    \immediate\write18{texcount -sub=section \jobname.tex  | grep "Section" | sed -e 's/+.*//' | sed -n \thesection p > 'count.txt'}
(\input{count.txt}words)}

\usepackage[english]{babel}

\usepackage[a4paper,top=2.5cm,bottom=2.5cm,left=2.5cm,right=2.5cm,marginparwidth=1.75cm]{geometry}

\usepackage{amsmath}
\usepackage{graphicx}
\usepackage[dvipsnames]{xcolor}
\usepackage{SIunits}
\usepackage{float}
\usepackage[export]{adjustbox}
\usepackage[colorlinks=true, allcolors=blue]{hyperref}
\DeclareUnicodeCharacter{2212}{-}
\usepackage{nameref}
\usepackage{pdfpages}
\usepackage{setspace}
\usepackage{subcaption}
\usepackage{booktabs} % For professional looking tables
\usepackage{multirow}
\usepackage{comment}
\usepackage{multicol}
\usepackage{longtable}
\usepackage{tabularx}
\usepackage{multirow}
\usepackage{algorithm2e}
\usepackage[english]{babel}
\usepackage[square,numbers]{natbib}
\begin{document}
\nocite{*}

\title{How reproducible are data-driven subtypes of Alzheimer’s disease atrophy?}

\author{Emma Prévot $^{a, b}$  \\ \texttt{emma.prevot@exeter.ox.ac.uk}\\
$^{a}$ {\footnotesize Department of Medical Physics and Biomedical Engineering, University College London,}\vspace{-5mm} \\  {\footnotesize Gower Street, London, WC1E 6BT, United Kingdom}  \vspace{-2mm} \\
$^{b}$ {\footnotesize  Department of Statistics, University of Oxford,}\vspace{-5mm} \\  {\footnotesize 24-29 St Giles', Oxford OX1 3LB, United Kingdom}  \vspace{-5mm}\\
\and Cameron Shand $^{c, d}$  \\ \texttt{cameron.shand@crick.ac.uk} \\
$^{c}$ {\footnotesize UCL Centre for Medical Image Computing, Department of Computer Science,  } \vspace{-5mm} \\ {\footnotesize University College London, Gower Street, London, WC1E 6BT, United Kingdom} \vspace{-2mm} \\
$^{d}$ {\footnotesize Software Engineering \& AI STP, The Francis Crick Institute, London, United Kingdom} \vspace{-5mm} \\
\and for Alzheimer’s Disease Neuroimaging Initiative $^{e}$  \\
$^{e}$ {\footnotesize A complete listing of ADNI investigators can be found at \href{http://adni.loni.usc.edu/wp-content/uploads/how_to_apply/ADNI_Acknowledgement_List.pdf}{this link}} \vspace{-5mm} \\ 
\and Neil P.~Oxtoby $^{c}$ \\ \texttt{n.oxtoby@ucl.ac.uk}\\
$^{c}$ {\footnotesize UCL Centre for Medical Image Computing, Department of Computer Science,  } \vspace{-5mm} \\ {\footnotesize University College London, Gower Street, London, WC1E 6BT, United Kingdom} \vspace{-2mm} \\
}

\date{}

\maketitle

\newpage

\newpage

\section*{ABSTRACT} 
The heterogeneity of Alzheimer’s disease (AD) is a major confound to treatment development and deployment. While new computational methods offer insights into AD progression, many lack validation. This study aims to assess the reproducibility of the AD progression subtypes identified by the Subtype and Stage Inference (SuStaIn) algorithm across distinct and independent databases. 

SuStaIn was employed on T1w MRI volumes from 5444 individuals across four cohorts constructed from the ANMerge, OASIS, and ADNI databases.  Eight models were created, with half using the full cohort and half excluding the control population to assess subtype robustness.

All three original SuStaIn AD subtypes, namely Typical, Cortical, and Subcortical, emerged from all cohorts. Additionally, the algorithm also characterised rare and atypical AD manifestations, such as posterior cortical atrophy (PCA).

This study validated the three main atrophy subtypes of AD across four datasets using Subtype and Stage Inference. This reproducibility bodes well for deploying such inference in clinical applications from stratification for interventional trials to differential diagnosis, but improving ethnic diversity of common databases remains a priority. \\ \\
\textbf{Keywords: }Dementia, heterogeneity, brain regions, disease progression, unsupervised machine learning.

\newpage

\section{INTRODUCTION}
Alzheimer’s disease (AD) stands as the predominant age-related neurodegenerative disorder and principal cause of dementia. Affecting 60-80\% of the estimated 55 million global dementia cases \cite{Dementia_stats, WHO_stats}, AD poses a significant burden with healthcare costs expected to increase to more than \$1 trillion by 2050, in light of an aging population \cite{Econ_stats}. However, despite being first described more than a century ago, disease-modifying therapies are only just now beginning to emerge --- and not without controversy. With over 200 clinical trials failed in the last 15 years \cite{Fail_Trial}, it is clear that the understanding of the causes of AD is limited. A significant obstacle to understanding and treating AD is the vast heterogeneity in clinical, genetic and pathophysiological biomarkers. Recent work has revealed the importance of distinguishing between temporal heterogeneity (different disease stage/severity) and phenotypic heterogeneity (different pathophysiological cascades) \cite{ferreira2020biological, murray2011neuropathologically,zheng2021molecular} to provide a comprehensive picture of the disease and enable accurate patient stratification and recruitment in clinical trials.

Within this complex landscape, the proliferation of publicly available datasets including demographic, clinical, and biologic information of individuals, has catalysed innovative insights into AD's pathophysiology, harnessing computational data-driven disease progression modelling and big data \cite{OxtobyAlexander}. We focus our attention to the Subtype and Stage Inference (SuStaIn) algorithm \cite{ZScoreSustain,pysustain}, an unsupervised machine-learning technique which combines disease progression modelling and clustering to perform subtyping and staging of individuals. Unlike traditional data-driven models that either assume a universal temporal progression \cite{EBM1,DPmodelNoH1,DPmodelNoH2}, or subtype while neglecting temporal insights \cite{ClusteringAlone2, PostMortem3, PostMortem2, ClusteringAlone1,murray2011neuropathologically}, SuStaIn uniquely disentangles various heterogeneity levels, stratifying patients that are temporally and phenotypically heterogeneous, based on a wide range of disease biomarkers. In the context of our study, we deploy SuStaIn on volumetric data across several brain regions to infer multiple subtype atrophy patterns. 

Previous analyses of AD atrophy subtypes using SuStaIn have revealed three distinct sequences of temporal progression \cite{ZScoreSustain}, which will be referred to as the original subtypes throughout. These clusters exhibit phenotypic similarities to patterns of atrophy revealed by post-mortem histology \cite{PostMortem1} and retrospective analysis of MRI scans close to the time of death \cite{SuStaInsub2, SuStaInsub1,structuredDatabase1}. However, in the original Z-score study, SuStain was exclusively applied to either synthetic data or data from the Alzheimer’s Disease Neuroimaging Initiative (ADNI) \cite{ADNI}, which is a well-characterised research dataset with a significant imbalance in terms of its ethnic representation \cite{structuredDatabase2, structuredDatabase1}. One exception is a follow-up study from Archetti et al.~\cite{studyExpanded} which focused on 
demonstrating the transferability of the SuStaIn disease progression models \cite{ZScoreSustain}. They trained SuStaIn on ADNI and tested the algorithm on a heterogeneous and less structured cohort, using data from three additional independent and less well-phenotyped datasets. Despite successfully demonstrating reproducibility on ``lower-quality'' MRI data, we identified some limitations. First, echoing their discussion, the number of individuals used to validate SuStaIn was limited and the brain regions selected were different compared to the original study. We also observe a lack of emphasis on biological relevance in this context. Finally, we believe the exclusive focus on testing the algorithm on less-structured datasets, rather than training on such datasets, may not fully address the challenges of adapting to ``lower-quality'' data as it does not consider the algorithm's ability to capture the intricacies and variations present in these datasets. \\

In this work, we extend the investigation of the transferability of the SuStaIn Alzheimer’s disease progression model by training and validating the algorithm on a larger cohort, and ensuring consistency between the biomarkers used. The aim is to assesses the reproducibility of the original subtypes and the reliability of the algorithm as a patient stratification tool, which is a necessary step before introducing it in the clinical environment. Furthermore, to the best of our knowledge, this study represents the first attempt to explore the impact of excluding healthy subjects (controls) from the model on disease progression subtypes. Additionally, we take a step further by demonstrating SuStaIn can characterise very specific atrophies, potentially aiding in the identification of less typical patterns of AD such as posterior cortical atrophy (PCA).

\newpage

\section{MATERIALS AND METHODS}
We begin by providing a high-level experimental overview that comprises key elements of our approach. First, we obtained the data from the chosen datasets and divided participants into four cohorts. We then established a control group, which played a crucial role as the reference for z-scoring and aided in adjusting for key confounding factors. Subsequently, we trained atrophy subtype models for each cohort using SuStaIn and determined the optimal number of clusters per model using 10-fold cross-validation. This was repeated both with the control group integrated into the model fitting process and without. Finally, we qualitatively and quantitatively analysed the resulting eight cross-validated models and their disease progression subtypes. 

\subsection{Participants and cohorts}
\begin{figure}[H]
\centering
\includegraphics[width=\textwidth]{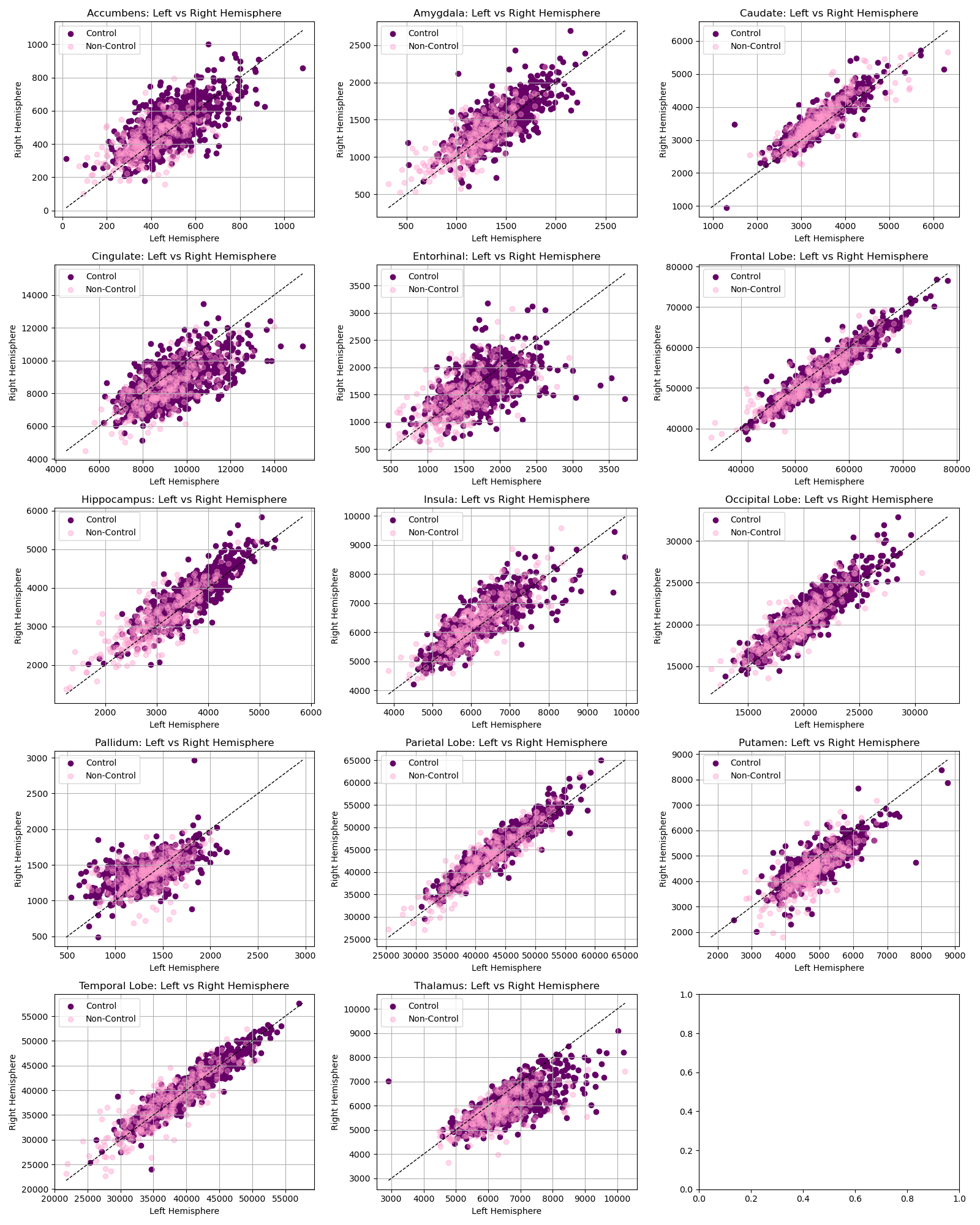}
\caption{\small \textbf{Scatter plots of left vs. right hemisphere ROIs measurements for the OASIS cohort} The only region in which we observe a slight imbalance is the Thalamus. However, this region was not originally included in the SuStaIn study and is not characteristic of any subtype. }
\label{fig:oasis_hemisphere_comparison_oasis}
\end{figure}

\begin{figure}[H]
\centering
\includegraphics[width=\textwidth]{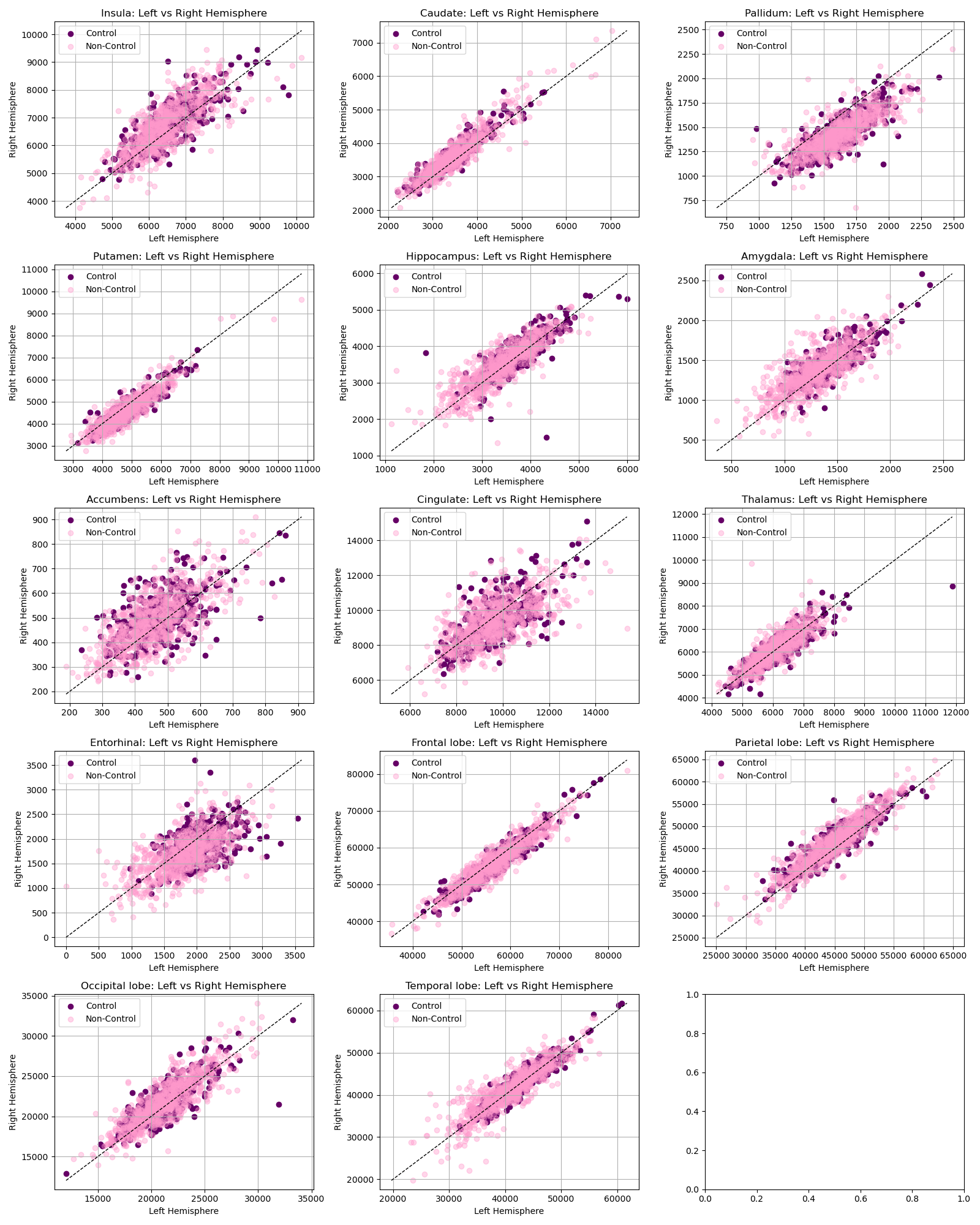}
\caption{\small \textbf{Scatter plots of left vs. right hemisphere ROIs measurements for the 3.0T ADNI cohort using only ROIs brain volumes} We don't observe any imbalance. }
\label{fig:hemisphere_comparison_adni_vol}
\end{figure}

\begin{figure}[H]
\centering
\includegraphics[width=\textwidth]{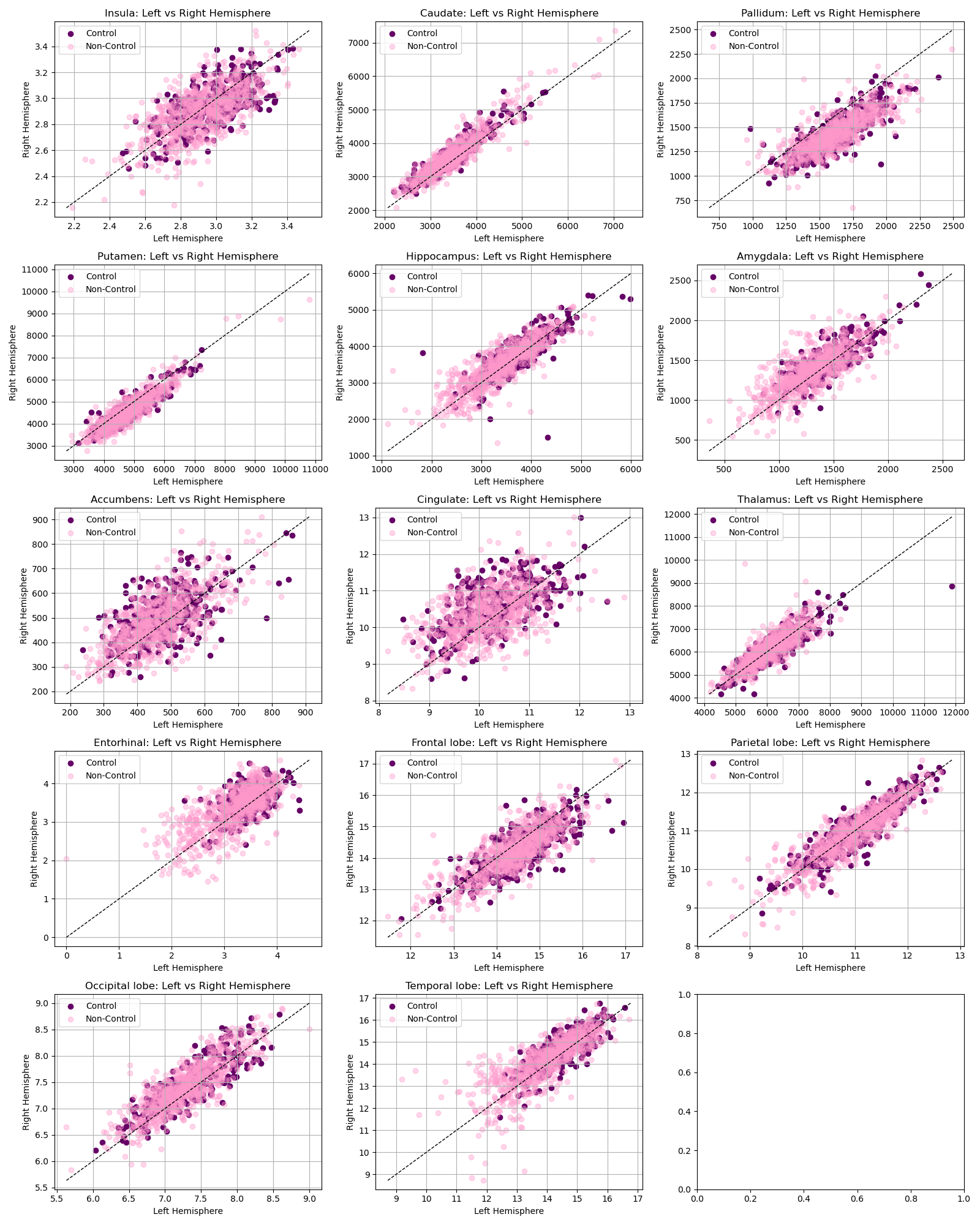}
\caption{\small \textbf{Scatter plots of left vs. right hemisphere ROIs measurements for the 1.5T ADNI cohort using ROIs brain volumes and thicknesses.} We don't observe any imbalance. }
\label{fig:hemisphere_comparison_adni_thick}
\end{figure}

 SuStaIn model fitting was performed with cross-sectional volumetric MRI data from 3 different databases: the Alzheimer’s Disease Neuroimaging Initiative (ADNI) \cite{ADNI}, the Open Access Series of Imaging Studies (OASIS) \cite{OASIS}, and the new version of the AddNeuroMed dataset (ANMerge) \cite{ANMERGE}. 
 Extracted data included demographic and clinical assessments, apolipoprotein E (APOE) genotype, intracranial volume, and numerical regional brain volumes or cortical thicknesses of 14 brain regions.  These regions of interest (ROIs), which are used to construct the disease progression sequences, include the Hippocampus, Amygdala, Nucleus Accumbens, Insula, Cingulate, Caudate, Pallidum, Putamen, Thalamus, Entorhinal cortex, and the Frontal, Temporal, Occipital, and Parietal lobes. We chose these regions following FreeSurfer lobe mapping as they were also used in the original SuStaIn study to allow direct comparison of findings. For ADNI and OASIS datasets, we extracted volumes for all ROIs. In the case of ANMerge, due to the unavailability of volumetric data for some regions, we extracted cortical thickness measurements for the Insula, Cingulate, Entorhinal Cortex, Frontal, Temporal, Occipital, and Parietal lobes, and volumes for the remaining regions. To combine ADNI with ANMerge, we also extracted cortical thickness measurements for the same ROIs in ADNI. In the Supplementary Material we detail the experiments conducted on the ADNI dataset to confirm that the SuStaIn model fitted using cortical thickness is equivalent to the one based on volumes, ensuring that no significant bias was introduced. However, being aware that cortical thinning is one of the earliest detectable signs of cognitive decline, we anticipate that the regions where thickness data were used might show earlier signs of atrophy \cite{pacheco2015greater}.  We averaged the volumes and thicknesses of the left and right hemispheres for a unified representation.  This was done after confirming that there were no significant asymmetries in the atrophy patterns between the two hemispheres in the averaged ROIs, for all cohorts, as shown in Supplementary Figures \ref{fig:hemisphere_comparison_diagnosis_anmerge},  \ref{fig:oasis_hemisphere_comparison_oasis}, \ref{fig:hemisphere_comparison_adni_thick} and \ref{fig:hemisphere_comparison_adni_vol}.   Inclusion was based on data availability, and only participants with complete entries for the extracted data were included, using baseline covariate data to maintain cross-sectionality.

ADNI data was downloaded from LONI's Imaging Data Archive (IDA) and two independent data sets were constructed: one including 3T (field strength) MRI data ($n=352$) and the other 1.5T MRI data ($n=1153$), both pre-processed with FreeSurfer version 5.1 to obtain cross-sectional volumes. ANMerge data was downloaded from the Synapse data portal and 1.5T MRI data pre-processed with FreeSurfer version 5.3 was downloaded ($n=931$). For both ADNI and ANMerge the individuals were broadly diagnosed as either Cognitively Normal (CN), Mild Cognitive Impairments (MCI), or Alzheimer's Disease (AD). OASIS data was downloaded from XNAT Central, a publicly accessible data repository, and 3T MRI data pre-processed with FreeSurfer version 5.3 was extracted ($n=1038$). The diagnosis for each individual was not presented in a straightforward categorical manner as in the other datasets, and instead, it contained additional extraneous and potentially confusing information.

These 3 databases were used to construct 4 different cohorts to investigate atrophy subtype reproducibility across T1w MRI magnetic field strength and USA/Europe: ANMerge ($n=931$), OASIS ($n=1038$), combined ADNI1.5T/ANMerge ($n=2084$) and combined ADNI3T/OASIS ($n=1391$). Both the ANMerge and OASIS MRI data acquisitions were designed to be compatible and comparable with ADNI \cite{ANMERGE, OASIS}. A primary goal of this work is to investigate whether differences in field strength, which can affect estimated volumes and thicknesses of brain regions, influence the reproducibility of atrophy subtypes. As for the different FreeSurfer versions, a check of the release notes confirmed no significant differences affecting the regions used in modeling between the FreeSurfer versions.

\subsection{Data preparation}

\subsubsection{Control population}

The definition of the control population was based on the Clinical Dementia Rating (CDR) scale. This metric evaluates cognitive and behavioural performances to stage the severity of dementia. Six different domains are assessed: memory, orientation, judgment and problem solving, community affairs, every-day life activities, and personal care \cite{CDR1,CDR2}. A subject with a CDR score of zero can be considered cognitively normal, i.e. not affected by any form of dementia. Thus, the control population of each cohort was identified with subjects having $\text{CDR}=0$. Table \ref{DEMcohort} shows the size of each control population with respect to that of the full cohort, as well as the percentage of carriers of at least one APOE $\epsilon 4$ allele for each group, which has been long associated with AD susceptibility \cite{apoe}, and The Mini Mental State Examination (MMSE) result for each group, which is a widely used test to screen for Dementia \cite{galasko1990mini}. A perfect MMSE score is 30, while a score of 24 is the recommended cutpoint for Dementia \cite{gluhm2013cognitive}.

Despite being a widely-used scale, CDR relies on cognitive tests, which are not as reliable as other typical Alzheimer’s cerebrospinal fluid (CSF) biomarkers such as $\beta$-amyloid \cite{CDRreliability}, especially for early diagnosis or individuals at preclinical stages. Nonetheless, such information was not available across all databases used in this study. Clinical diagnoses were also unsuitable due to inconsistencies in classification between ADNI, ANMerge, and OASIS, with the latter providing extraneous information alongside diagnoses, hampering its reliability and comparability as a control separator. Therefore, to maintain consistency across cohorts, CDR was chosen to define the control population. As shown in Table \ref{DEMcohort}, the $CDR = 0$ control groups align well with our expectations for cognitively normal individuals, characterised by near-perfect MMSE scores and a lower prevalence of APOE $\epsilon4$ allele carriers. Moreover, in the Supplementary Material we describe experiments we carried on the ADNI dataset to confirm that using solely $CDR=0$ as the control separator, or combining it with amyloid-negativity, as was done in the original SuStaIn study \cite{ZScoreSustain}, leaves the identified subtypes unchanged.

\begin{table}[H]
\centering
\setlength{\tabcolsep}{7pt}
\resizebox{\textwidth}{!}{% use resizebox with textwidth
\begin{tabular}{llllllll}
\toprule
Cohort &  & $n$ & APOE  & Sex & Age & MMSE & CDR \\
& & & (\% of $\epsilon 4$) & (\% of F) & (yrs, s.d.) & (mean, s.d.) & (mean, s.d.) \\
\midrule
ANMerge & Controls & 273 &  30\%  & 54\% &  74.3 $\pm$ 6.2  & 29.0 $\pm$ 1.2 & - \\
 & Patients & 658 &  53\%  & 60\% &  75.0 $\pm$ 6.5  &  23.2 $\pm$ 4.8 & 1.02 $\pm$ 0.69 \\
 \midrule
OASIS & Controls & 736 &  32\%  & 60\% &  67.4 $\pm$ 8.2 &   29.1 $\pm$ 1.3 & -  \\
 & Patients & 302 & 51\% & 40\% & 72.6 $\pm$ 7.6  &  25.3 $\pm$ 4.6  & 0.72 $\pm$ 0.42 \\
 \midrule
A1.5/A & Controls & 610 &  29\% & 53\% & 73.6 $\pm$ 6.1 &  29.0 $\pm$ 1.2 &  - \\
 & Patients & 1474 &  51\%  & 52\% &  73.4 $\pm$  7.4 &  25.4 $\pm$ 4.3 & 0.70 $\pm$ 0.36 \\
  \midrule
A3/O  & Controls & 849 & 34\%  & 60\% & 67.9 $\pm$ 9.4 &  29.1 $\pm$ 1.2 &  - \\
 & Patients & 542 &  55\% & 44\%  &  73.4 $\pm$ 7.8 & 25.6 $\pm$ 3.8 & 0.65 $\pm$ 0.33 \\
\bottomrule
\bottomrule
\end{tabular}
}
\caption{\small \label{DEMcohort} \textbf{Demographic, clinical, and genetic information of the four constructed cohorts, divided between Controls ($CDR=0$), and Patients ($CDR>0$). } \\\textit{Abbreviations: APOE $\epsilon 4$ --- percentage of carriers of at least one APOE $\epsilon 4$ allele; $n$ --- number of subjects in the cohort; M --- Male; F --- Female; MMSE --- Mini Mental State Examination; CDR --- Clinical Dementia Rating.}
}
\end{table}

\subsubsection{Detrending and Z-scoring}

Covariate correction (detrending) was performed to regress out the effect of age, sex, and intracranial volume (ICV) on the brain region measurements \cite{covariate1,covariate2}, given the inter-cohort demographic variability. Subsequently, data was z-scored relative to the control population; each regional brain volume was expressed as a z-score by subtracting the mean of the control population and dividing by the control standard deviation. By doing so, z-scored regional volumes and thicknesses indicate how many standard deviations away from the control mean (i.e. normality) each subject value is, which is indicative of abnormality. Given that brain volumes reduce with AD progression, leading to decreasing and eventually negative z-scores, we flip the sign of the calculated z-scores allowing them to increase with advancing atrophy, as expected by the SuStaIn algorithm.

While we acknowledge that W-scores, which directly adjust for covariates, are a more commonly used approach in similar studies \cite{iaccarino2021spatial, jack1997medial, katsumi2023association}, we opted to use Z-scores after performing separate covariate detrending. This decision was made to remain consistent with the original SuStaIn study methodology, allowing more direct comparison of findings.

\subsection{SuStaIn model}

SuStaIn, or the Subtype and Stage Inference algorithm \cite{ZScoreSustain}, is a data-driven disease progression model designed to identify unique disease progression subgroups within populations by unraveling their phenotypic and temporal heterogeneity. It can employ multiple event-based disease progression models \cite{EBM1} including a linear z-score model \cite{ZScoreSustain}. The Z-score model characterises disease progression as a linear increase in ROIs abnormalities, measured in z-scores, i.e., standard deviations from the mean of the healthy (control) subjects. SuStaIn can work with purely cross-sectional data, creating a common timeframe divided into stages for regional atrophies evolution. We direct the reader to the original reference for more details \cite{ZScoreSustain}.

SuStaIn performs hierarchical clustering to group subjects with similar disease progression patterns into subtypes. It starts with all data in one cluster, estimating the disease sequence, then progressively subdivides into more subtypes, each time recalculating the sequence. For each cluster, SuStaIn estimates the proportion of subjects, the most probable z-score sequence for each ROI, and assigns each subject to the most probable stage in the sequence. The relative likelihood of each sequence is approximated evaluating the probability of a number of possible sequences sampled using Markov Chain Monte Carlo (MCMC) \cite{MCMC}, which also gives a visual and quantitative measure of the uncertainty associated with each subtype z-score ordering.

Two models were constructed for each cohort, one incorporating and the other excluding the control population in the model fitting. This yields a total of eight experiments, numbered as follows: models 1 and 2 were constructed on ANMerge 1.5T MRI data, with and without controls respectively, and similarly models 3 and 4 for the combined ADNI and ANMerge 1.5T MRI data, 5 and 6 for OASIS 3T MRI data, and finally 7 and 8 for the combined ADNI and OASIS 3T MRI data. Removing control subjects from the model fitting is an element of novelty of our study. The assumption is that if the control group is accurately defined, its inclusion or exclusion should not significantly impact SuStaIn’s subtyping and staging, as by model specification we would expect almost all the controls to be below $z==2$ in any given ROI.

\subsubsection{pySuStaIn settings}
The SuStaIn algorithm is publicly available through the pySuStaIn software package available at \url{https://github.com/ucl-pond}.

To construct a SuStaIn model, several parameters are established, but this discussion focuses on those altered from the original Z-score study \cite{ZScoreSustain}. The z-score event thresholds used, i.e. the number of standard deviations away from the mean of the healthy controls which can be considered an abnormal event, were only 2 and 3, omitting z-score 1 to avoid erroneous conclusions on the severity of the atrophy. A z-score 2 event is the linear accumulation from any z-score up to z-score 2, which corresponds to a brain region reaching a value that is 2 standard deviations away from the control mean. Similarly, the z-score 3 event occurs when a region z-score linearly accumulates from z-score 2 to 3.

\subsubsection{Cross-Validation}
As SuStaIn proceeds hierarchically, when running it with $N_{max}$ maximum subtypes, it will also construct the ($N_{max}-1$) subtypes models.  Across each SuStaIn model (for each number of subtypes up to the pre-defined maximum) 10 $k$-fold cross-validation is used to capture model uncertainty. The test set log-likelihood and the Cross Validation Information Criterion (CVIC) \cite{CVIC} are then used for model selection, including choosing the best number of clusters, aiming for consistent log-likelihood increases and lower CVIC for optimised model complexity and accuracy balance.

The number of maximum clusters per model was initially set to 4, with adjustments made after inspecting output visualisations, test set log-likelihood and CVIC to chose the best subtype model.

\subsection{Subtype and Stage Analysis}

For each subtype in each model, SuStaIn outputs one Positional Variance Diagram (PVD) which summarises the inferred sequence of disease progression and its uncertainty. Each sequence is divided into 28 stages (14 ROIs $\times$ 2 z-score events). Additionally, individuals with no abnormality (according to the model) in all 14 examined brain regions are allocated to Stage 0.

In the original Z-score study \cite{ZScoreSustain}, three neuroanatomical AD subtypes, namely the Typical, Cortical and Subcortical, were found when modelling data from the Alzheimer’s Disease Neuroimaging Initiative (ADNI). SuStaIn revealed that for the Typical subtype, atrophy started in the Amygdala and Hippocampus; for the Cortical subtype, it began in the Cingulate, Insula, and Nucleus Accumbens; for the Subcortical subtype in the Caudate, Pallidum, Putamen, and Nucleus Accumbens. These three clusters emerged from both 1.5T and 3T MRI data.

We qualitatively assessed whether one subtype could be identified as either Typical, Cortical, or Subcortical \cite{ZScoreSustain} looking at the regions exhibiting early severe atrophy in the PVDs. Subsequently, we also asses late-stage atrophy observations to provide a richer characterisation. When comparing subtypes between different models the number of cross-validated subtypes and the MRI field strength (either 1.5T or 3T) were also considered, the latter specifically to account for potential variances in smaller brain regions.

\subsubsection{Cognitive assessments analysis}
For specific atrophy patterns, we explored cognitive assessment scores of the corresponding individuals. This was primarily performed for the ANMerge cohort, where a broad collection of neurocognitive and psychological assessments are available, including single question or single task scores, which we divided into different domains as shown in Supplementary Table \ref{ANMergecogass}. The following tests were analysed: the Clinial Dementia Rating (CDR), the Mini Mental State Examination (MMSE) score; the Alzheimer’s Disease Assessment Scale-Cognitive Subscale (ADAS-Cog); the Geriatric Depression Scale (GDS); and the Alzheimer’s Disease Cooperative Study Activities of Daily Living Scale (ADCS-ADL). In the OASIS database, only summative scores of CDR and MMSE were available. For combined cohorts this investigation was not possible due to inconsistencies in available data and protocols across databases.

\subsubsection{Statistical analysis}
We statistically assessed the similarity and differences between subtypes across cohorts. Pairwise Kendall's tau distances of the ROIs z-score sequences was used to quantify cross-cohort similarity of the Typical, Cortical, and Subcortical subtypes. Qualitatively, a Kendall rank correlation coefficient greater than 0.50 suggests strong similarity, while a coefficient below 0.30 suggesting weak similarity \cite{schober2018correlation}. 

Additional model and subtype comparisons included ANOVA, two-proportions z-tests, and pairwise t-tests with Bonferroni correction (for the number of subtypes) across demographic, cognitive, genetic, and CSF features. ANOVA was used to compare means among subtypes, such as age or MMSE score. For a more targeted pairwise comparison, ensuring we account for multiple comparisons and control the family-wise error rate, we used pairwise t-tests coupled with a Bonferroni correction. Finally, when it came to evaluating proportional differences, especially when the data was categorical, the two-proportion z-test was used.

\newpage 

\section{RESULTS}
Table \ref{similarity} provides an overview of all the models from the 8 cohorts, constructed from selected combinations of data from ADNI, ANMerge, and OASIS: 1.5T, 3.0T, with/without controls during model fitting. Overall, the three original subtypes of SuStaIn discovered in ADNI data \cite{ZScoreSustain} were replicated across most cohorts. Additionally, the cross-validated models also included mixed clusters with multiple patterns of atrophy, clusters made of outliers with no significant atrophy pattern, and also a subtype exhibiting strong and early atrophy in the posterior cortices which resembled posterior cortical atrophy (PCA).   The exclusion of control data in model fitting often resulted in the Subcortical and/or Cortical subtype disappearing, suggesting the need for longitudinal validation of the subtypes. In the subsequent sections, we elaborate on each of them individually. For clarity, we will use  ``clusters" to refer to subgroups identified by the algorithm, and ``subtypes" when these groups have been assigned a clinical interpretation.

\begin{table}[H]
\centering
\resizebox{\textwidth}{!}{% use resizebox with textwidth
\begin{tabular}{l|l|cccc|cccc}
\hline \multicolumn{2}{c}{ } & \multicolumn{4}{|c|}{ \textbf{WITH CONTROLS} } & \multicolumn{4}{c}{ \textbf{WITHOUT CONTROLS} } \\
\toprule
\hline \begin{tabular}{l} 
\textbf{Cohort}
\end{tabular} & \textbf{MFS} & \textbf{X} &  \textbf{Subtypes} & \textbf{N}  & \textbf{CVIC} & \textbf{X}  & \textbf{Subtypes} & \textbf{N} & \textbf{CVIC} \\
\hline
\begin{tabular}{l} 
ANMerge 
\end{tabular} & 1.5 T & 1 &  \begin{tabular}{c} 
\textcolor{red}{\textbf{Typical}}\\
\textcolor{blue}{\textbf{Cortical}} \\
PCA subtype
\end{tabular} & 3  & 42869 & 2  & \begin{tabular}{c} 
\textcolor{red}{\textbf{Typical}} \\
\textcolor{blue}{\textbf{Cortical}} \\
PCA subtype
\end{tabular} & 3 & 30288\\
\hline
\begin{tabular}{l} 
ANMerge + \\
ADNI 1.5T
\end{tabular} & 1.5 T & 3  &  \begin{tabular}{c} 
\textcolor{red}{\textbf{Typical}} \\
\textcolor{blue}{\textbf{Cortical}} \\
\textcolor{olive}{\textbf{Subcortical}} \\
Mixed
\end{tabular} & 4 & 94147 & 4  & \begin{tabular}{c} 
\textcolor{red}{\textbf{Typical}} \\
\textcolor{blue}{\textbf{Cortical}} \\
Mixed
\end{tabular} & 3 & 66318\\
\hline
\begin{tabular}{l} 
OASIS 
\end{tabular} & 3.0 T & 5   & \begin{tabular}{c} 
\textcolor{red}{\textbf{Typical}} \\
\textcolor{blue}{\textbf{Cortical}} \\
\textcolor{olive}{\textbf{Subcortical}}
\end{tabular} & 3 & 46413 & 6  &  \begin{tabular}{c} 
\textcolor{red}{\textbf{Typical}} \\
Outliers
\end{tabular} & 2 & 12990\\
\hline
\begin{tabular}{l} 
OASIS + \\
ADNI 3.0T
\end{tabular} & 3.0 T & 7  & \begin{tabular}{c} 
\textcolor{red}{\textbf{Typical}} \\
\textcolor{blue}{\textbf{Cortical}} \\
\textcolor{olive}{\textbf{Subcortical}}
\end{tabular}& 3 & 63778 & 8  &  \begin{tabular}{c} 
\textcolor{red}{\textbf{Typical}} \\
Outliers
\end{tabular} & 2 & 25043\\
\bottomrule
\bottomrule
\end{tabular}

}

\caption{\small \label{similarity} 
\textbf{Summary of the SuStaIn subtypes in the constructed models.} The color-coding is to help visualise the three original subtypes from \cite{ZScoreSustain} --- Typical, Cortical, Subcortical --- reproduced across the eight experiments. Additionally, we refer to \textit{Mixed} subtype when several atrophy patterns were identified within the same disease progression sequence; \textit{Outliers} instead indicates a subtype with no recognisable atrophy pattern. The CVIC column reports the optimum value found for the corresponding number of subtypes $N$ in the model. \\ \textit{Abbreviations: MFS --- Magnetic Field Strength; X --- Experiment number; $N$ --- Number of cross-validated subtypes; CVIC --- Cross Validation Information Criterion.}
} 
\end{table}

\subsection{SuStaIn atrophy models}

\subsubsection{1.5T atrophy models, ANMerge dataset}

\begin{figure}[H]
\centering
\includegraphics[width=1\textwidth]{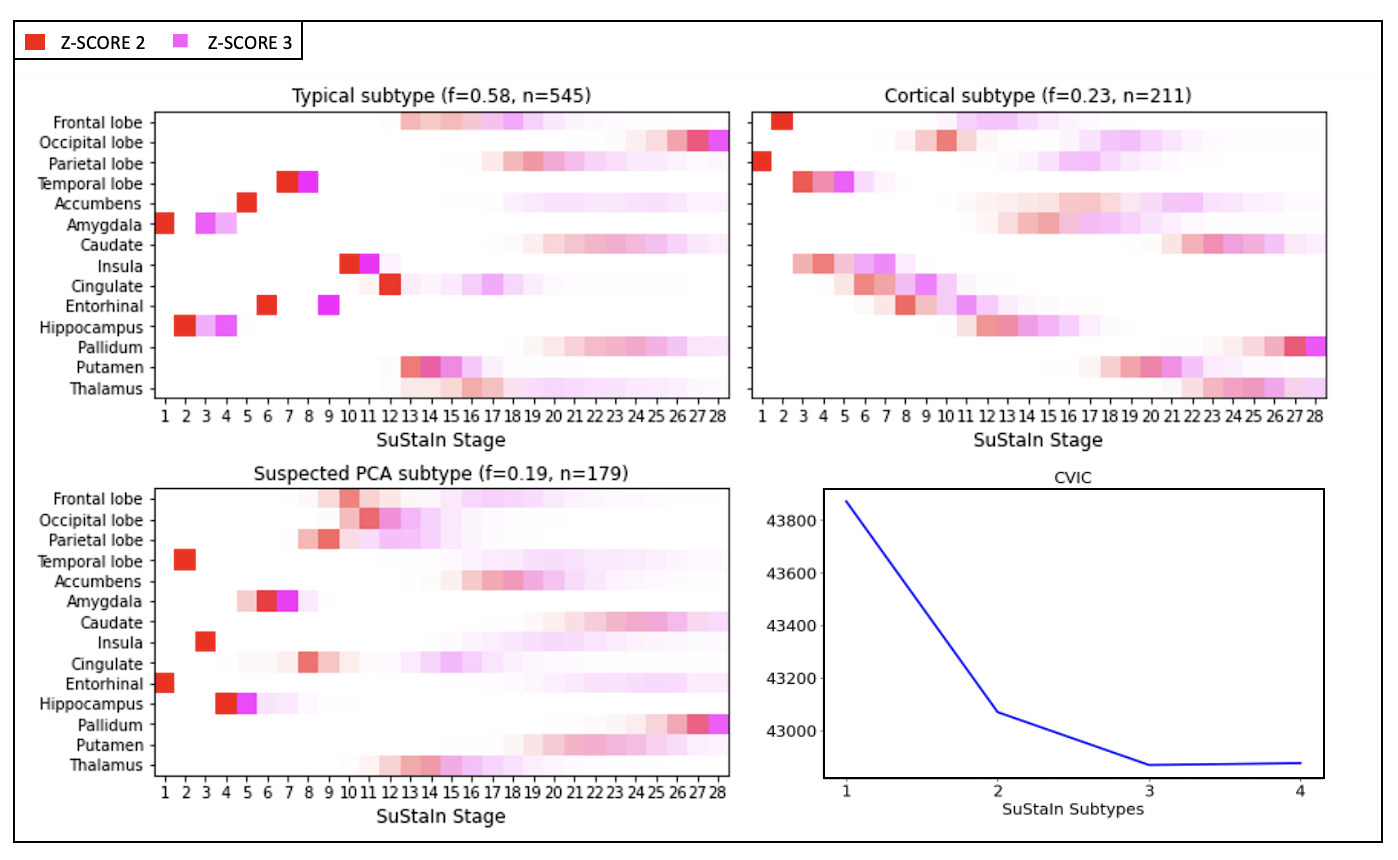}
\caption{\small \label{fig:anmerge}
\textbf{Results of experiment 1: 3-subtype atrophy model in 1.5T MRI, CN+MCI+AD participants, ANMerge dataset.} The lower right panel shows that optimum CVIC occurs for $N=3$ subtypes. The remaining panels are positional variance diagrams \cite{EBM1} (PVDs), where the vertical axes show the 14 input features (regional brain volumes). Each progression pattern consists of a sequence of stages in which the corresponding regional brain volume reaches different z-scores relative to the control population, in a probabilistic sequence (left-to-right). At each stage the colour in each region indicates the level of severity of volume loss: white is unaffected; red is moderately affected (z-score of 2); magenta is severely affected (z-score of 3).  $f$ and $n$ indicate the average fraction and number of individuals assigned to each subtype across MCMC samples. The uncertainty of each event in the sequence can be inferred by the blurriness of the PVD.\\\textit{Abbreviations: PCA --- posterior cortical atrophy; CVIC --- Cross Validation Information Criterion; CN --- Cognitively Normal; MCI --- Mild Cognitive Impairments; AD --- Alzheimer's Disease.}
}
\end{figure}

Figure \ref{fig:anmerge} shows positional variance diagrams \cite{EBM1} for the cross-validated 3-subtype atrophy progression pattern estimated from 1.5T MRI in CN+MCI+AD participants from the ANMerge study, along with the CVIC model comparison plot (lower right panel). We can recognise the Typical subtype with severe and early atrophy affecting the Hippocampus and Amygdala, and the Cortical subtype with atrophy starting in the Insula and Cingulate. The third subtype, which we recognised as the Suspected PCA subtype, initially resembles typical AD atrophy, followed by notably earlier atrophy in the posterior cortices. 

Statistical comparison of demographic and cognitive outcomes across these subtypes revealed that nearly 27\% of the subjects in the suspected PCA subtype were 65 years old or younger, which is significantly higher than 6\% and 2\% for the Typical and Cortical subtype respectively (\textit{two-proportions z-test}, $p\ll0.05$). Additionally, comparing the mean age of individuals with pairwise t-tests with a Bonferroni correction suggests that subjects assigned to the suspected PCA subtype are younger than the typical and cortical ones ($p \approx 0.03$). In agreement with clinical knowledge on parieto-occipital atrophy, measures of visual capabilities (extracted from MMSE and ADAS-Cog) are worse for these individuals (\textit{two-proportions z-test}, $p<0.05$) --- more than 80\% of parieto-occiptal subtype individuals reproduced a drawing shown to them incorrectly, and more than 60\% were only able to name zero or one object when shown to them.

The Subcortical subtype found in previous analyses of 3T MRI data \cite{ZScoreSustain} was undetected in the 1.5T ANMerge data; however, after adjusting the model to include the z-score 1 event threshold, as per the original z-score study, nearly 30\% of subjects exhibited early z-score 1 atrophy in subcortical regions.

Supplementary Figure \ref{fig:anmergeNC} shows the three cross-validated clusters when controls were not included in model fitting. The same three atrophy subtypes were produced, including an even earlier atrophy in the Occipital and Parietal lobes for the suspected PCA subtype. Again, individuals assigned to this subtype were younger on average than those assigned to the typical and/or cortical subtypes (ANOVA test, $p \ll 0.05$). The suspected PCA subtype also displayed exacerbated difficulties in calculation, copying drawings, and naming objects that were shown to them  (\textit{two-proportions z-test}, $p < 0.001$ vs Cortical subtype and $p \ll 0.0001$ vs Typical subtype).

\subsubsection{1.5T atrophy models, ADNI and ANMerge combined}

\begin{figure}[H]
\centering
\includegraphics[width=1\textwidth]{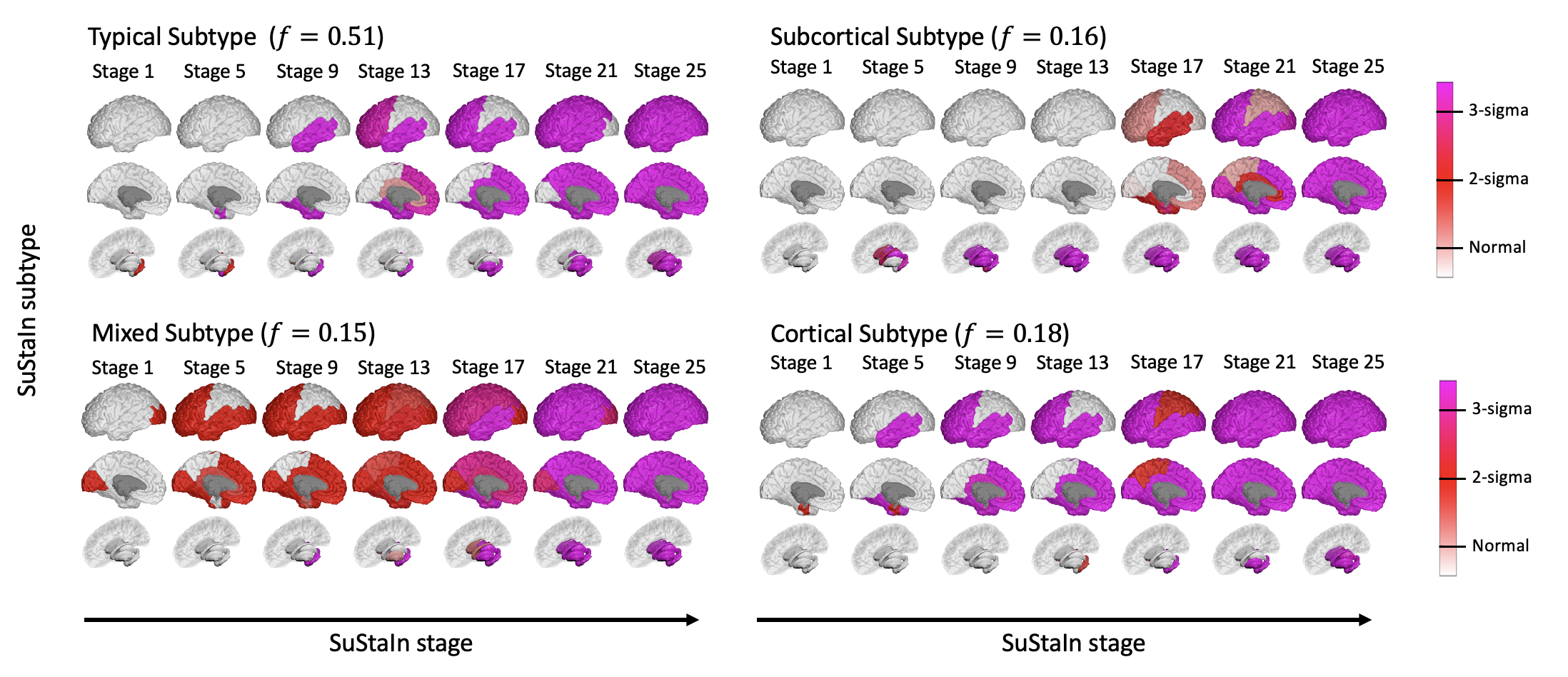} 
\caption{\small \label{fig:adnianmergeC}
\textbf{Results of experiment 3: 4-subtype atrophy model in 1.5T MRI, ADNI and ANMerge dataset.}  The progression pattern of each of the four subtypes that SuStaIn identifies \cite{marinescu2019brain}. Each progression pattern consists of a sequence of stages in which regional brain volumes reach different z-scores relative to the control population. At each stage the colour in each region indicates the level of severity of volume loss: white is unaffected; red is moderately affected (z-score of 2); magenta is severely affected (z-score of 3). $f$ is the proportion of participants estimated to belong to each subtype.
\\\textit{Abbreviations:  CVIC --- Cross Validation Information Criterion; PVD --- Positional Variance Diagram}
}
\end{figure}

Figure \ref{fig:adnianmergeC} shows the resulting PVD for the model trained on the combined 1.5T MRI data from ADNI and ANMerge. Four subtypes were cross-validated, three of which broadly replicate the original findings in \cite{ZScoreSustain} --- Typical, Cortical, Subcortical --- with addition of a mixed Typical/Cortical subtype. Beyond the expected atrophies from the original SuStaIn model \cite{ZScoreSustain}, the Typical subtype exhibited frontal lobe abnormality, the Cortical subtype included fronto-temporal abnormalities, and the Subcortical subtype also involved early abnormality in the Thalamus, Hippocampus, and Amygdala. The fourth subtype showed a typical/cortical pattern of atrophies, but no lobe was affected by severe shrinking ($z==3$ events).
Supplementary Figure \ref{fig:adnianmergeNC} shows the three cross-validated subtypes without controls in the model fitting. The clusters are similar to the disease sequences constructed with controls, excluding the Subcortical subtype which was not found. 

\subsubsection{3.0T atrophy models, OASIS dataset}

 Supplementary Figure \ref{fig:oasisC} shows the results of experiments on 3T MRI data from the OASIS study. Broadly speaking, the three subtypes replicated the original findings in \cite{ZScoreSustain} --- Typical, Cortical, Subcortical --- but with high positional variance/uncertainty (blurriness). Notable differences with the original findings include earlier involvement of the Frontal lobe in the Typical subtype, and early severe atrophy of the Caudate in the Subcortical subtype. Without controls in the model fitting, cross-validation suggested only two subtypes but the difference was marginal and fitting a 3-subtype model without controls revealed the Subcortical subtype once again with remarkably abnormal atrophies in Caudate and Putamen (Supplementary Figure \ref{fig:oasisNC}).

\subsubsection{3.0T atrophy models, ADNI and OASIS combined}

\begin{figure}[H]
\centering
\includegraphics[width=0.9\textwidth]{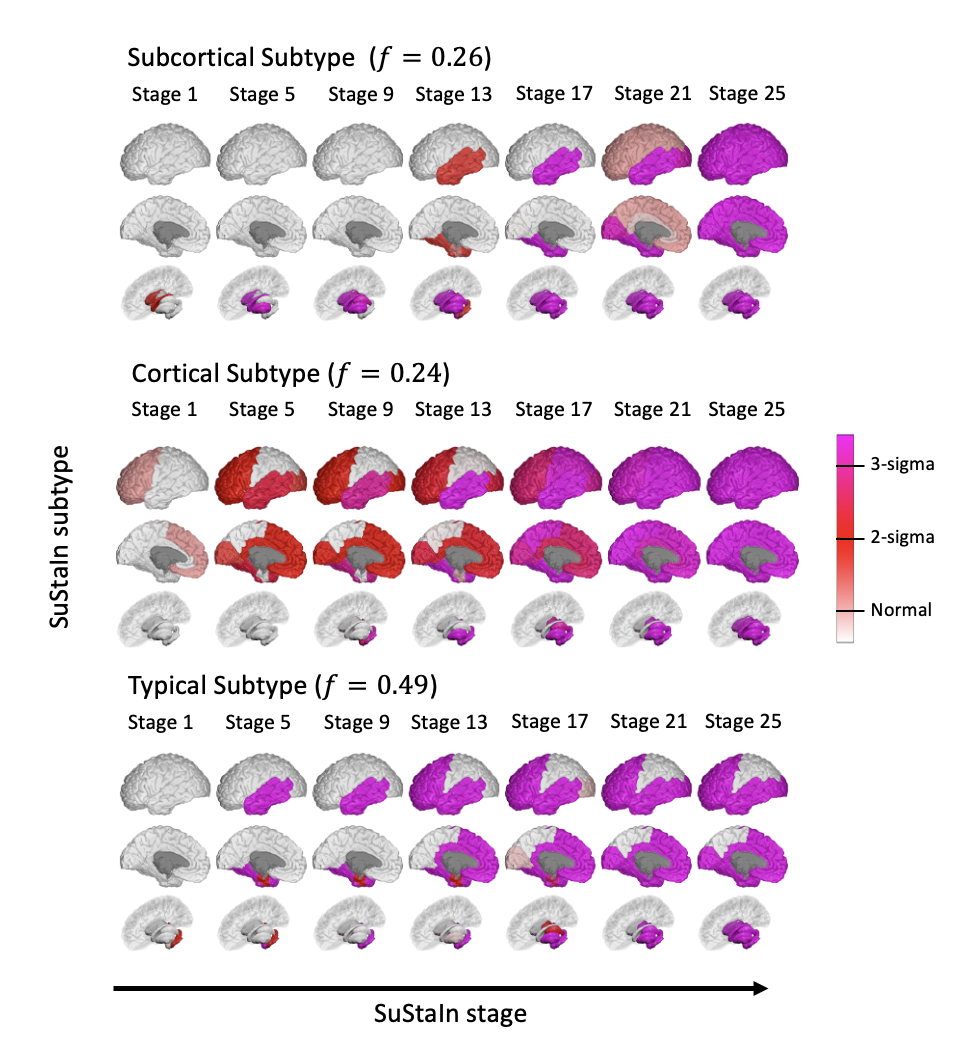}
\caption{\small \label{fig:adnioasis}
\textbf{Results of experiment 7: 3-subtype atrophy model in 3.0T MRI, ADNI and OASIS dataset.}  The progression pattern of each of the three subtypes that SuStaIn identifies \cite{marinescu2019brain}. Each progression pattern consists of a sequence of stages in which regional brain volumes reach different z-scores relative to the control population. At each stage the colour in each region indicates the level of severity of volume loss: white is unaffected; red is moderately affected (z-score of 2); magenta is severely affected (z-score of 3). $f$ is the proportion of participants estimated to belong to each subtype.}
\end{figure}

Figure \ref{fig:adnioasis} shows the results of experiments on 3.0T MRI data from ADNI and OASIS studies. Cross-validation supported a 3-subtype model, which broadly reproduced the three original clusters in \cite{ZScoreSustain} --- Typical, Cortical, Subcortical. Notably, the Typical pattern additionally exhibited early and severe frontal atrophy, and the Subcortical pattern featured early involvement of the Caudate. Supplementary Figure \ref{fig:adnioasisNC2} shows the resulting subtypes with controls excluded from model fitting. Cross-validation supported a 2-subtype model having high positional variance/uncertainty (blurriness), with more than 95\% of the cohort staged earlier than stage 9, leaving only data from a small number of individuals from which to estimate the latter parts of each subtype sequence. Only the Typical subtype was clearly identified; the second subtype exhibited mild atrophy (z-score 2) in various regions and a severe (z-score 3) atrophy in the Caudate. 

\subsection{Similarity between subtypes across models}

Table \ref{similarity} shows that the original subtypes from \cite{ZScoreSustain} were successfully replicated in distinct datasets. Supplementary Figures \ref{fig:typical tau}, \ref{fig:cortical tau}, and \ref{fig:subcortical tau} summarise the pairwise Kendall's tau correlation coefficients for the atrophy sequences of the Typical, Cortical, and Subcortical subtypes, respectively. Agreement between model subtype atrophy sequences was generally high, especially within database and within cohort (with/without controls). The Typical subtype emerged across all eight models and it consistently manifested late atrophies in cortical regions, demonstrating the highest level of concordance in the progression sequence across all cohorts (Kendall's tau correlation between $0.53$ and $1.0$). The Cortical subtype appeared in six of eight models and the often progressed into typical atrophies, except for the OASIS cohorts and ADNI3T/OASIS cohorts, where Putamen was also involved around similar stages. The pairwise Kendall's tau correlation coefficients between the cortical subtypes ranged between $0.36$ and $0.99$, indicating strong, or moderately strong similarity. The lowest concordance was evident between models using entirely different data sets and contrasting approaches to the inclusion or exclusion of controls, such as between the combined OASIS and ADNI3.0T model and the combined ANMerge and ADNI1.5T model without controls.   The Subcortical subtype emerged only in three of eight models and was notably impacted by the removal of control subjects, the magnetic field strength of the MRI scans, and the decision to omit the z-score 1 event. As for the Cortical subtype, this subtype was often accompanied by typical atrophies around similar stages. The pairwise Kendall's tau correlation coefficients between the subcortical subtypes across different models ranged between $0.51$ and $0.98$, indicating very strong similarity.

Table \ref{statSimDiff} summarises demographic, clinical, biological and genetic variables across subtypes in each cohort, highlighting statistical differences or similarities. In all cohorts the proportion of control subjects participating in the disease progression sequence (i.e., stage $> 0$) is significantly higher in the Cortical and Subcortical subtypes but the proportion of subject genetically more susceptible to AD, i.e., carrying at least one APOE $\epsilon 4$ allele, is significantly higher in the Typical subtype. We do not consistently observe significant difference in demographic variables like age and sex\footnote{It is worth noting that we did not include in Table \ref{statSimDiff} the posterior cortical atrophy subtype found in ANMerge data.}. Importantly, despite the imbalance in the proportion of controls and the genetic AD susceptibility, we did not observe consistent significant differences in cognition as measured by the MMSE.

Supplementary Figures \ref{fig:ternarycomb} and \ref{fig:ternary} show the probability of subtype assignment for each subject, indicative of SuStaIn's confidence, across each cohort, separated into controls and non-controls. Subjects are generally strongly assigned to the Typical subtype, as expected. Additionally, subtype assignment was considerably less confident in the OASIS data, especially controls.

\begin{table}[H]
\centering
\setlength{\tabcolsep}{7pt}
\resizebox{\textwidth}{!}{% use resizebox with textwidth
\begin{tabular}{lllllllll}
\toprule
%Cohort  & Subtype & $n$ &  Control & APOE $\epsilon 4$ & Sex (M/F) & Age (yrs, s.d.) & MMSE (mean, s.d.) & Stage (mean, s.d.) \\

Cohort & Subtype & $n$ & Control & APOE $\epsilon 4$ & Sex & Age & MMSE & Stage \\
& & & & & (\% of F) & (yrs, s.d.) & (mean, s.d.) & (mean, s.d.) \\

\midrule
ANMerge & Typical & 319 & 4\% $^{a}$ & 66\% $^{a}$ & 69\% $^{a}$ & 75.7 $\pm$ 6.3 $^{a}$ & 22.9 $\pm$ 4.9  & 4.4 $\pm$ 3.9 $^{a}$\\
 & Cortical & 123 &  22\% $^{a}$ & 41\% $^{a}$ & 54\% $^{a}$ & 74.3 $\pm$ 5.4 $^{a}$ & 23.9 $\pm$ 5.8 & 5.8 $\pm$ 5.2 $^{a}$\\
 & Subcortical & - & -  & - & - & - & - & -\\
 \midrule
OASIS & Typical & 151 &  34\% $^{a, b}$ & 58\% $^{a}$ & 52\% & 71.5 $\pm$ 3.4 &  25.1 $\pm$ 4.4 $^{a, b}$ & 3.4 $\pm$ 3.4 $^{a, b}$\\
 & Cortical & 133 &  58\% $^{a, c}$ & 38\% $^{a}$ & 57\% & 69.5 $\pm$ 1.8 & 27.1 $\pm$ 3.6 $^{a, c}$ & 2.4 $\pm$ 1.9 $^{a    }$\\
 & Subcortical & 54 &  72\% $^{b, c}$ & 50\% & 50\% & 69.2 $\pm$ 1.6 & 28.3 $\pm$ 2.4 $^{b, c}$ & 1.9 $\pm$ 1.6 $^{b    }$\\
 \midrule
A1.5/A & Typical & 613 &  4\% $^{a, b}$ & 61\% $^{a, b}$ & 55\% & 74.4 $\pm$ 6.9 $^{a, b}$ & 24.2 $\pm$ 4.4 & 4.0 $\pm$ 3.4 \\
 & Cortical & 185 &  22\% $^{a}$ & 31\% $^{a, c}$ & 58\% & 77.0 $\pm$ 5.5 $^{a, c}$ & 24.0 $\pm$ 5.5 & 4.5 $\pm$ 4.5\\
 & Subcortical & 170 &  26\% $^{b}$& 47\% $^{b, c}$ & 50\% & 71.0 $\pm$ 7.5 $^{b, c}$ & 25.1 $\pm$ 4.6 & 4.1 $\pm$ 3.6\\
  \midrule
A3/O  & Typical & 301 &  21\% $^{a, b}$ & 56\% $^{a}$ & 53\% $^{b}$ & 72.3 $\pm$ 8.7 & 25.9 $\pm$ 3.1& 2.8 $\pm$ 2.6 $^{b}$\\
 & Cortical & 134 &  58\% $^{a}$  & 38\% $^{a}$ & 57\% $^{c}$ & 71.0 $\pm$ 8.1 & 28.1 $\pm$ 2.4& 2.4 $\pm$ 1.9\\
 & Subcortical & 57 &  60\% $^{b}$ & 47\% & 37\% $^{b, c}$ & 72.2 $\pm$ 7.7 & 27.9 $\pm$ 2.7& 1.9 $\pm$ 1.6 $^{b}$\\
\bottomrule
\bottomrule
\end{tabular}
}
\caption{\small \label{statSimDiff} \textbf{Descriptive statistics of demographic, genetic and cognitive variables for individuals assigned to each subtype in each cohort when including controls in model fitting.} We removed the subjects assigned to Stage 0 as those are not participating in the disease progression sequence. Values marked with $^{a}$ indicate a significant difference ($p<0.05$) between Typical and Cortical subtype, $^{b}$ between Typical and Subcortical, and $^{c}$ between Cortical and Subcortical. \\\textit{Abbreviations: $n$ --- number of subject assigned to the subtype; Control --- percentage of subject assigned to the subtype with $CDR=0$; APOE $\epsilon 4$ --- percentage of subject assigned to the subtype with at least one $\epsilon 4$ allele; M --- Male; F --- Female; MMSE --- Mini Mental State Examination} 
}
\end{table}

\newpage 

\section{DISCUSSION}
This study was designed to investigate the reproducibility of Alzheimer’s disease atrophy subtypes identified in the original SuStaIn study \cite{ZScoreSustain}, as a function of MRI scanner field strength, data source (study), and whether controls were included in the model fitting. 
Our results show that the three original atrophy subtypes discovered by SuStaIn --- Typical, Cortical and Subcortical --- consistently emerged across datasets and scanner field strength, with some variation depending on whether controls were included in the model fitting. 

\begin{figure}[H]
\centering
\includegraphics[width=1\textwidth]{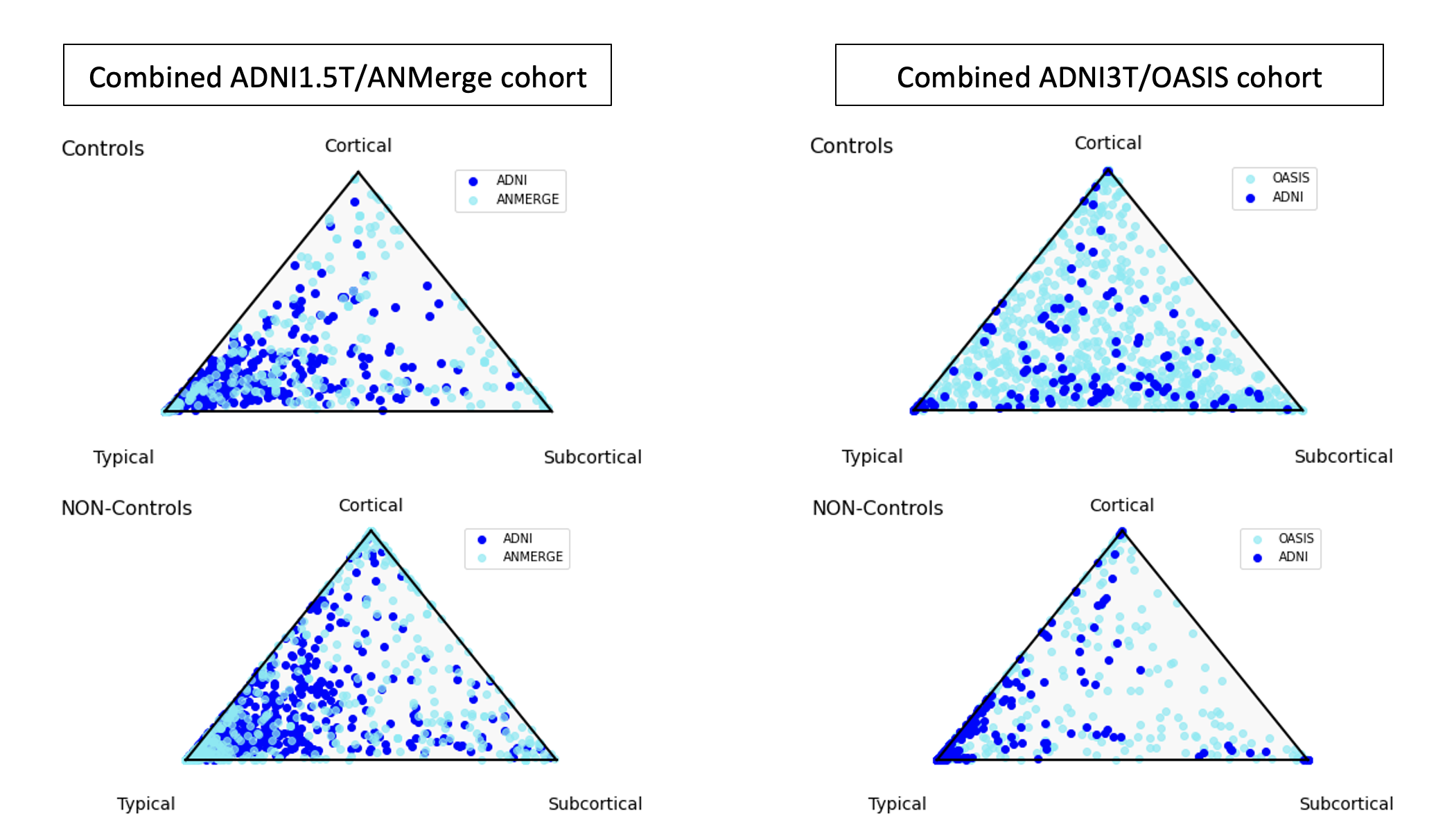}
\caption{\small \label{fig:powders}\textbf{Probability of subtype assignment for the combined cohorts.} The closest an individual (dot) is to a triangle vertix, the higher the probability of assignment to the corresponding subtype. Subjects are generally strongly assigned to the Typical subtype. However, this is not the case for the OASIS control population, which behaves differently from the non-control individuals and has several subjects clustered in the centre of the triangle. This suggests the high level of uncertainty associated with the subtyping of these subjects. }
\label{fig:ternarycomb}
\end{figure}

\begin{figure}[H]
\centering
\includegraphics[width=1\textwidth]{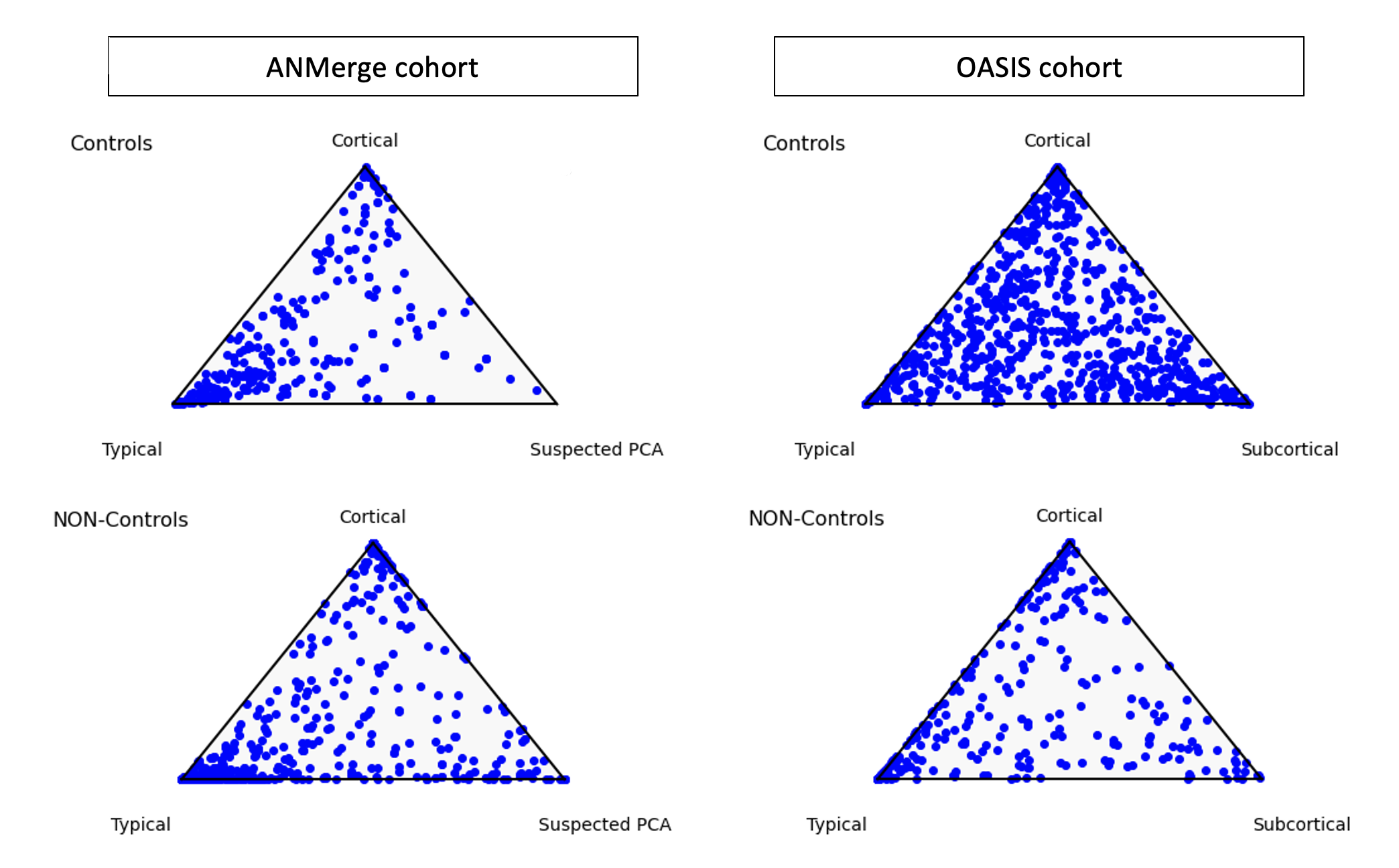}
\caption{\small \label{fig:powders}\textbf{Probability of subtype assignment for the single cohorts.}  Similar to Fig.\ref{fig:ternarycomb}, the OASIS controls are equally distributed across the subtypes and there is a high level of uncertainty, demonstrated by the number of subjects clustered in the middle of the triangle. Remarkably, very few control subjects were assigned to the suspected PCA subtype in ANMerge.}
\label{fig:ternary}
\end{figure}

The Typical subtype, characterised by early atrophy in the hippocampus and amygdala, aligns with the classic AD pattern often described in literature \cite{Braak, ZScoreSustain}. This was the largest subtype across all datasets (Table \ref{statSimDiff}) and demonstrated the highest level of assignment consistency and confidence across cohort (Supplementary Figure \ref{fig:ternarycomb} and \ref{fig:ternary}). Notably, the Typical subtype was always assigned the smallest fraction of control subjects and the highest fraction of APOE $\epsilon 4$ allele carriers (Table \ref{statSimDiff}), consistent with a canonical manifestation of AD. In terms of agreement across cohorts, both the presence of the Typical subtypes and their internal consistency within cohorts were noteworthy. Kendall’s tau calculations (Supplementary Figure \ref{fig:typical tau}), indicated a high degree of intra-subtype inter-cohort similarity, with almost perfect sequence concordance between the same cohort including or excluding the control population.

The Subcortical subtype presented a more complex picture. While it generally showed similarity with the corresponding cluster defined in the original SuStaIn work, it also paralleled the hippocampal-sparing variant of AD described in prior research \cite{Similarsubtypes, PostMortem1}, partly explaining why it continually included the highest representation of control subjects. Moreover, it was characterised by a higher prevalence of male subjects (Table \ref{statSimDiff}), which is in line with previous research \cite{studyExpanded, ferreira2020biological}. Nonetheless, an anomaly was observed whereby the ANMerge data did not support a Subcortical subtype under cross-validation (although this was a borderline result and model comparison using CVIC, or any other criterion, is not perfect). There are two possible explanations for this. First, the limitations of 1.5T MRI scans may have reduced the signal to noise ratio in the small volumetric variations in subcortical regions. Secondly, model hyperparameters have a significant impact --- when adding a subtle atrophy event of $z=1$ (as in the original SuStaIn analysis \cite{ZScoreSustain}), the Subcortical subtype returned. This implies that subcortical atrophy in AD variants is more subtle than in other affected regions. Across cohorts, Kendall’s tau (Supplementary Figure \ref{fig:subcortical tau}) suggests a high degree of intra-subtype inter-cohort similarity for the models supporting a Subcortical subtype. 

Finally, the Cortical subtype was found in most models and always mimicked the disease pattern found in the original study \cite{ZScoreSustain}, characterised by early insula and cingulate atrophy. The Cortical subtype consistently included the smallest fraction of APOE $\epsilon 4$ allele carriers, and a higher proportion of controls, similar to the Subcortical subtype (Table \ref{statSimDiff}). Notably in the OASIS models, the cortical pattern was not as evident (lower, albeit still moderately strong, intra-subtype inter-cohort similarity: Supplementary Figure \ref{fig:subcortical tau}) and seemed to be driven by data from controls --- when the control population was removed, the Cortical subtype disappeared. \\

The exclusion of controls data from model fitting had varying impacts on the three original atrophy subtypes --- Typical, Cortical and Subcortical.  Controls, by definition, should not exhibit AD-related atrophy, yet their inclusion in the model might provide insights into early-stage disease progression if they are in fact presymptomatic AD cases. The key concern, though, is whether the presence of controls could inadvertently incorporate the effects of normal aging into the AD subtypes, as it is well-documented that brain volumes, specifically cortical, decrease with age in cognitively unimpaired older adults \cite{cook2017rates, freeman2008preservation}. Our results suggest that the effect of including data from controls in AD progression model fitting is more complex than merely introducing normal aging patterns. We highlight one key observation per subtype. 
The Typical subtype was unaffected and appeared in all models. This is unsurprising as the subtype was driven mostly by patient data even when control data was included in model fitting, and all studies were enriched for the typical, memory-led clinical phenotype of AD. 
The Cortical subtype appeared consistently in 1.5T non-control data from ANMerge and ADNI, but not in the 3T non-control data that was predominantly from OASIS. This is plausibly explained by OASIS's relative enrichment for preclinical individuals \cite{OASIS}, with a substantial 71\% of participants having a CDR score of 0, combined with the relatively stringent minimum atrophy z-score of 2 in our modelling. Moreover, supplementary Figure \ref{fig:z-score dist cortical} demonstrates controls showed pronounced atrophies, underscoring the significance of their contribution to the sequence.
Finally, the Subcortical subtype disappeared altogether when control data was excluded from model fitting. There are multiple layers to this. First, the controls ($CDR=0$) apparently driving the Subcortical subtype, showed pronounced atrophies, with z-scores exceeding 2, akin to non-controls (Supplementary Figure \ref{fig:z-score dist}). Second, the Subcortical subtype might be expected to have an atypical clinical phenotype (at least early on), leading to under-representation among the non-control samples in these memory-enriched studies. This could render clustering analyses such as ours underpowered.
Focusing on the OASIS models, which were significantly impacted by the removal of control subjects, reveals one more insight.  Notably, 60\% of controls carried the $\epsilon 4$ allele of the APOE gene, linked to higher Alzheimer’s risk, significantly differing from the 20\% in other databases ($p \ll 0.05$). These individuals could be preclinical AD cases, thereby influencing the model outcomes when removed prior to fitting. 
Supported by our insights, we argue that including control subjects in AD models does more than merely introduce normal aging patterns. This is particularly pertinent for subcortical and cortical atrophy, which may not immediately affect memory and cognitive scores like the CDR \cite{gupta2019alzheimer}. However, the Cortical and Subcortical subtypes individuals would eventually manifest typical Alzheimer's-related atrophy and subsequent cognitive decline. If precision in defining control subjects is limited, their inclusion might still be beneficial for early AD pathology detection, outweighing risks of normal aging contamination. This is vital for early detection and intervention in AD, where neuropathological changes often precede cognitive symptoms by years \cite{Hallmarks2} and early-stage interventions, like anti-amyloid therapies, have shown increased effectiveness \cite{earlyAD,earlyAD2}. SuStaIn's ability to identify early AD signs thus becomes strategically important for AD treatment.

The analysis of the ANMerge cohort revealed a distinct subtype characterised by prominent atrophy in the occipital and parietal lobes, resonating with the neuropathology of Posterior Cortical Atrophy (PCA) \cite{PCA1,PCA2}. Consistent with the neurological damage, PCA manifests itself mainly with different forms of visual impairments, as well as difficulties in mathematical calculation and spelling. Moreover, PCA has a much earlier onset than normal AD as it commonly occurs between 50--65 years of age. Our findings align with these characteristics, as the suspected PCA cluster individuals performed significantly worse in tasks typically associated with PCA manifestations, with impairments becoming even more evident in non-control subjects.  Furthermore, the mean age of the suspected PCA cluster was significantly lower than the others, and, in the model not including the control population, the majority of the subjects (55\%) were 65 years old or younger. These findings support the hypothesis that the cluster found in ANMerge can be linked to PCA. SuStaIn's ability to discern this atypical subtype could prompt clinicians to conduct targeted clinical assessments alongside standard (memory-led) ones, which might fall short or misinterpret atypical AD and other neurodegenerative diseases \cite{cogNoAtypical}.

We highlight some strengths and limitations of our study. We assessed the SuStaIn model and AD models under different scenarios, thereby evaluating their reproducibility and robustness. Moreover, our methodology was further enhanced by incorporating a diverse array of neuropsychological assessments, where possible, to improve interpretability of our models. However, our study is not without limitations, which in turn, guide the trajectory for future research. While our models were validated across multiple datasets spanning two continents, it's important to contextualize our findings within the limited diversity of ethnicity and socioeconomic status represented in the data. This limitation, inherent to the public datasets employed in our research, underscores a broader challenge faced by many researchers in the field. Addressing this gap in future studies by expanding the ethnic diversity of cohorts will be crucial to validate and extend the applicability of our results \cite{ADNI4weiner2023increasing}.
Moreover, we have highlighted the CDR's shortcomings in capturing non-typical disease manifestation and progression. Disease progression models like the Z-score model must be interpreted with respect to the reference control population. However, the complexity and variability of AD make defining an ideal control group challenging \cite{SuStaInsub1}. Our basic definition (CDR=0) could be augmented in future work with biomarker data on beta-amyloid and/or phosphorylated-tau negativity \cite{p-tau2, p-tau3, p-tau1}, and genetic risk factors like the APOE4 allele.

In conclusion, this study uniquely reveals conditions under which one can expect robustness and inter-cohort reliability of the SuStaIn algorithm in its ability to reliably subtype and stage Alzheimer’s disease patients across distinct databases. The evident consistency in identifying three primary disease progression sequences --- Typical, Cortical, and Subcortical --- reinforces the notion that Alzheimer’s disease may in fact be a set of sub-diseases, or disease subtypes, rather than a single biological cascade.

\newpage 

\section{DATA AVAILABILITY}

ADNI data can be obtained via the ADNI and LONI websites (adni.loni.usc.edu). For up-to-date information, see www.adni-info.org. The OASIS-1: Cross-Sectional data can be obtained via http://www.oasis-brains.org.  AddNeuroMed data can instead be accessed through https://www.synapse.org. SuStaIn algorithm is available at https://github.com/ucl-pond.

\section{ACKNOWLEDGEMENTS}

The authors acknowledge members of the UCL POND group (\url{http://pond.cs.ucl.ac.uk}) for valuable feedback received during group discussions.
EP is supported by and acknowledges the Oxford EPSRC Centre for Doctoral Training in Health Data Science (EP/S02428X/1), which is currently funding her DPhil at the University of Oxford. \\
Part of the data used in this project was obtained from the Alzheimer’s Disease Neuroimaging Initiative (ADNI) database (\url{adni.loni.usc.edu}). The ADNI was launched in 2003 as a public-private partnership, led by Principal Investigator Michael W. Weiner, MD. The primary goal of ADNI has been to test whether serial magnetic resonance imaging (MRI), positron emission tomography (PET), other biological markers, and clinical and neuropsychological assessment can be combined to measure the progression of mild cognitive impairment (MCI) and early Alzheimer’s disease (AD). Data collection and sharing for the Alzheimer's Disease Neuroimaging Initiative (ADNI) is funded by the National Institute on Aging (National Institutes of Health Grant U19 AG024904), the Canadian Institutes of Health Research in Canada, and generous contributions from the following: AbbVie, Alzheimer’s Association; Alzheimer’s Drug Discovery Foundation; Araclon Biotech; BioClinica, Inc.; Biogen; Bristol-Myers Squibb Company; CereSpir, Inc.; Cogstate; Eisai Inc.; Elan Pharmaceuticals, Inc.; Eli Lilly and Company; EuroImmun; F. Hoffmann-La Roche Ltd and its affiliated company Genentech, Inc.; Fujirebio; GE Healthcare; IXICO Ltd.; Janssen Alzheimer Immunotherapy Research \& Development, LLC.; Johnson \& Johnson Pharmaceutical Research \& Development LLC.; Lumosity; Lundbeck; Merck \& Co., Inc.; Meso Scale Diagnostics, LLC.; NeuroRx Research; Neurotrack Technologies; Novartis Pharmaceuticals Corporation; Pfizer Inc.; Piramal Imaging; Servier; Takeda Pharmaceutical Company; and Transition Therapeutics. The grantee organization is the Northern California Institute for Research and Education. Private sector contributions were made possible through the Foundation for the National Institutes of Health (\url{http://www.fnih.org}).\\
Another part of the data was provided by OASIS-1: Cross-Sectional. The authors acknowledge the Principal Investigators: D. Marcus, R, Buckner, J, Csernansky J. Morris; P50 AG05681, P01 AG03991, P01 AG026276, R01 AG021910, P20 MH071616, U24 RR021382. 
Finally, data was also provided by AddNeuroMed. The authors acknowledge the AddNeuroMed project participants and the AddNeuroMed project scientists—clinical leads responsible for data collection —Iwona Kloszewska (Lodz), Simon Lovestone (London), Patrizia Mecocci (Perugia), Hilkka Soininen (Kuopio), Magda Tsolaki (Thessaloniki), and Bruno Vellas (Toulouse), imaging leads—Andy Simmons (London), Lars-Olof Wahlund (Stockholm) and Christian Spenger (Zurich), and bioinformatics leads—Richard Dobson (London) and Stephen Newhouse (London).

\section{FUNDING}
This work was supported by the UKRI Medical Research Council via both a Future Leaders Fellowship (MR/S03546X/1) and the Joint Programme -- Neurodegenerative Disease Research (E-DADS project: MR/T046422/1).

\section{COMPETING INTERESTS}
NPO is a consultant for Queen Square Analytics Limited (UK), including on initiatives related to data-driven subtyping of Alzheimer's disease. The authors declare no other relevant disclosures.

\clearpage

\setstretch{1.0}
%\printbibliography
\bibliography{bibliography}

\clearpage
\section*{SUPPLEMENTARY MATERIAL}

\subsection*{Control definition experiments}
We ran sanity checks on the ADNI cohort, which is the only one for which we could better define out definition of the control population. We ran three different SuStaIn models: one in which only $CDR=0$ was used to define the control population; one where we used amyloid-negative cognitively normal (CN) subjects as controls, defined as those with a CSF $A\beta1–42$ measurement $>192 pg$ per ml, as done in the original SuStaIn study \cite{ZScoreSustain}; and one where we used both $CDR=0$ and $A\beta1–42>192 pg$ per ml.

All individuals with $CDR=0$ are also amyloid-negative and diagnosed as cognitively normal. Therefore, the amyloid-negative CN individuals is exactly the same as the amyloid-negative CN individuals with $CDR=0$, therefore we only consider one of the two. We choose to compare the \textit{CDR only} control definition with \textit{CDR + Amyloid negativity} definition.  When using only CDR to separate control individuals, we obtain that 30\% of the ADNI cohort individuals are controls. When including amyloid-negativity, we exclude those $CDR=0$ individuals who are not amyloid negative and obtain that 20\% of the ADNI cohort individuals are controls. Therefore, the \textit{CDR only} control population is bigger than the \textit{CDR + Amyloid negativity} one.

\begin{figure}[H]
\centering
\includegraphics[width=\textwidth]{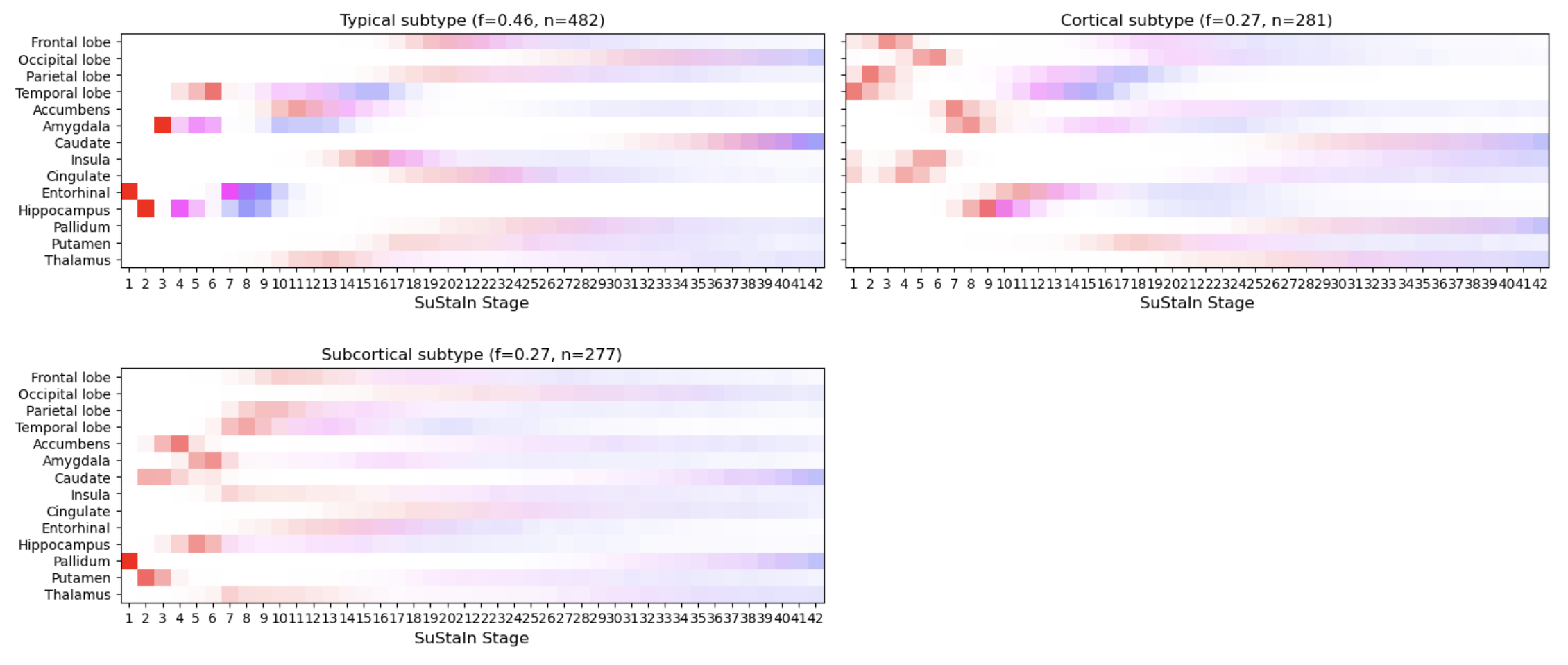}
\caption{\small \textbf{Results of experiment on ADNI using only CDR=0 as control separator.}  {\small The panels are PVDs \cite{EBM1} where the vertical axes shows the 14 regional brain volumes that reach atrophy z-scores along the x-axis of $z=1$ (red), $z=2$ (magenta) and $z=3$ (blue) in a probabilistic sequence (left-to-right). We recognise the three original SuStaIn subtypes \cite{ZScoreSustain}.}  \\\textit{Abbreviations: PVD --- Positional Variance Diagram; CDR --- Clinical Demential Rating.}}
\label{fig:justCDRvolumes}
\end{figure}

\begin{figure}[H]
\centering
\includegraphics[width=\textwidth]{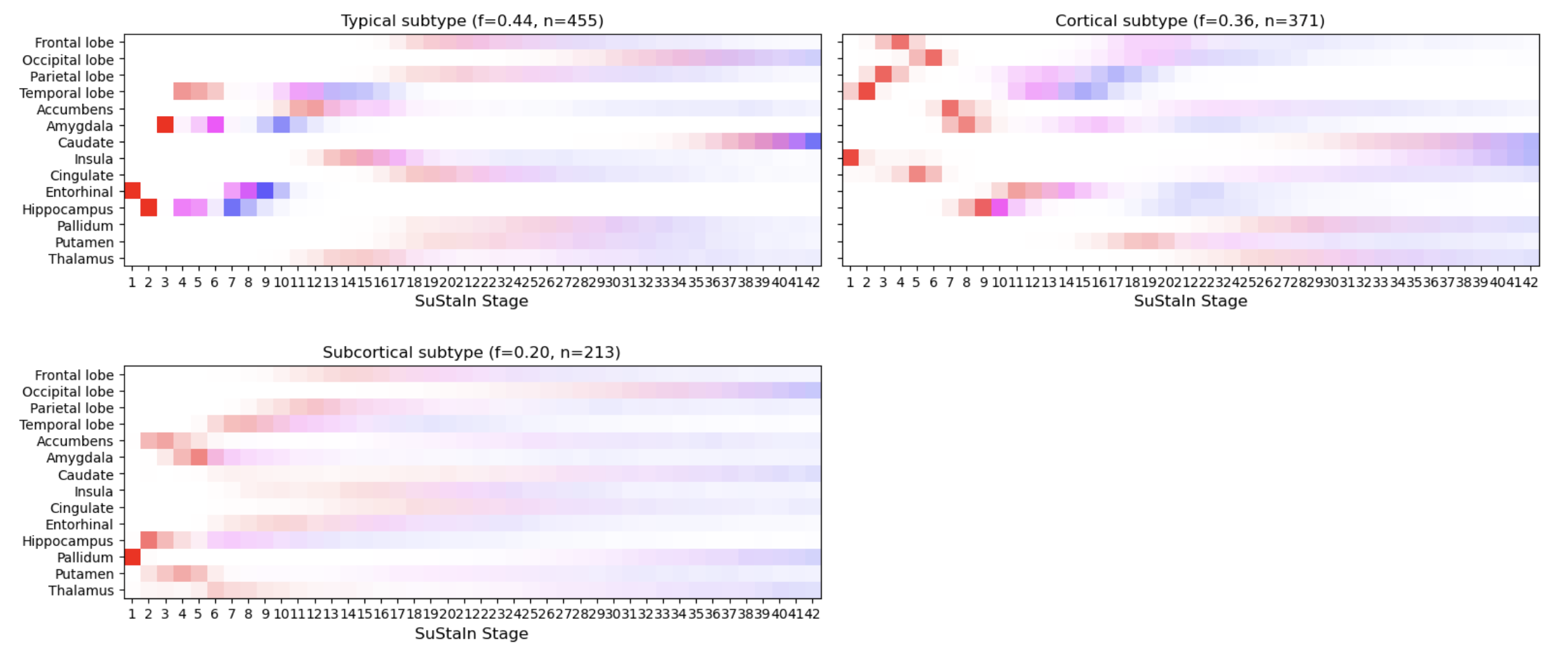}
\caption{\small \textbf{Results of experiment on ADNI using CDR=0 and Amyloid negativity as control separator.} The panels are PVDs \cite{EBM1} where the vertical axes show the 14 regional brain volumes that reach atrophy z-scores along the x-axis of $z=1$ (red), $z=2$ (magenta) and $z=3$ (blue) in a probabilistic sequence (left-to-right). We do still recognise the three original SuStaIn subtypes \cite{ZScoreSustain}.  \\\textit{Abbreviations: PVD --- Positional Variance Diagram; CDR --- Clinical Demential Rating.}}
\label{fig:CDR_amyloid}
\end{figure}

Figures \ref{fig:justCDRvolumes} and \ref{fig:CDR_amyloid} show the cross-validated SuStaIn models using the \textit{CDR only} and the \textit{CDR + Amyloid negativity} control definitions respectively. It is important to remember that changing the control definitions changes the reference point with respect to which data is z-scored, as different control individuals will lead to different averages for ROIs. However, the same subjects will be used to construct the models. All three subtypes are unchanged, some exhibit more uncertainty in some ROIs compared to others but overall the sequence of events is the same.

\begin{figure}[H]
\centering
\includegraphics[width=\textwidth]{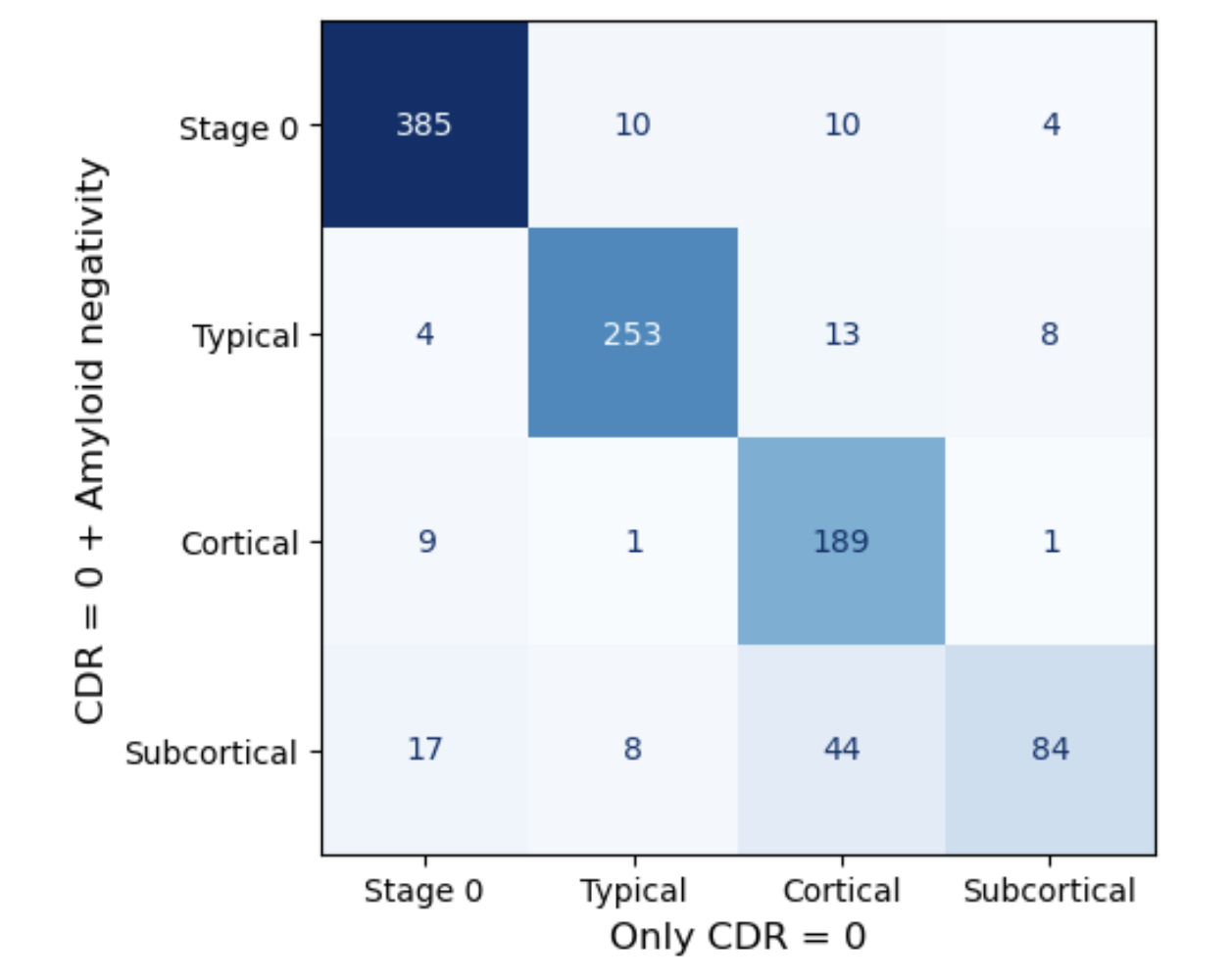}
\caption{\small \textbf{Confusion matrix of subtype assignment between the SuStaIn models fitted with two different control definitions.}  The x-axis indicates the $CDR=0$ control definition, the y-axis the $CDR=0$ + Amyloid negativity. Under Stage 0 we grouped all the individuals that were staged at zero, therefore exhibiting no sign of atrophy.  The numbers in the squares indicate the number of individuals assigned to each specific group.  \textit{Abbreviations: CDR --- Clinical Demential Rating}}
\label{fig:confusion_matric_cdr}
\end{figure}

Figure \ref{fig:confusion_matric_cdr} instead shows a confusion matrix of individuals assignment. We observe the vast majority of individuals are assigned to the same subtype, except little confusion between Subcortical and Cortical subtype.

\newpage

\subsection*{Cortical thickness vs. cortical volume experiments}
We ran sanity checks on the ADNI cohort, which is the cohort for which we extracted both cortical thickness and cortical volumes depending on whether we were combining it with ANMerge (thicknesses) or OASIS (volumes). The ROIs for which we extracted the thickness instead of the volume are Insula, Cingulate, Entorhinal Cortex, Frontal, Temporal, Occipital, and Parietal lobes. We fitted SuStaIn on either just cortical volumes for all 14 ROIs, or on cortical thicknesses for the regions we just mentioned, and volumes for all the other ones. 
We used $CDR=0$ as the control definition. 

\begin{figure}[H]
\centering
\includegraphics[width=\textwidth]{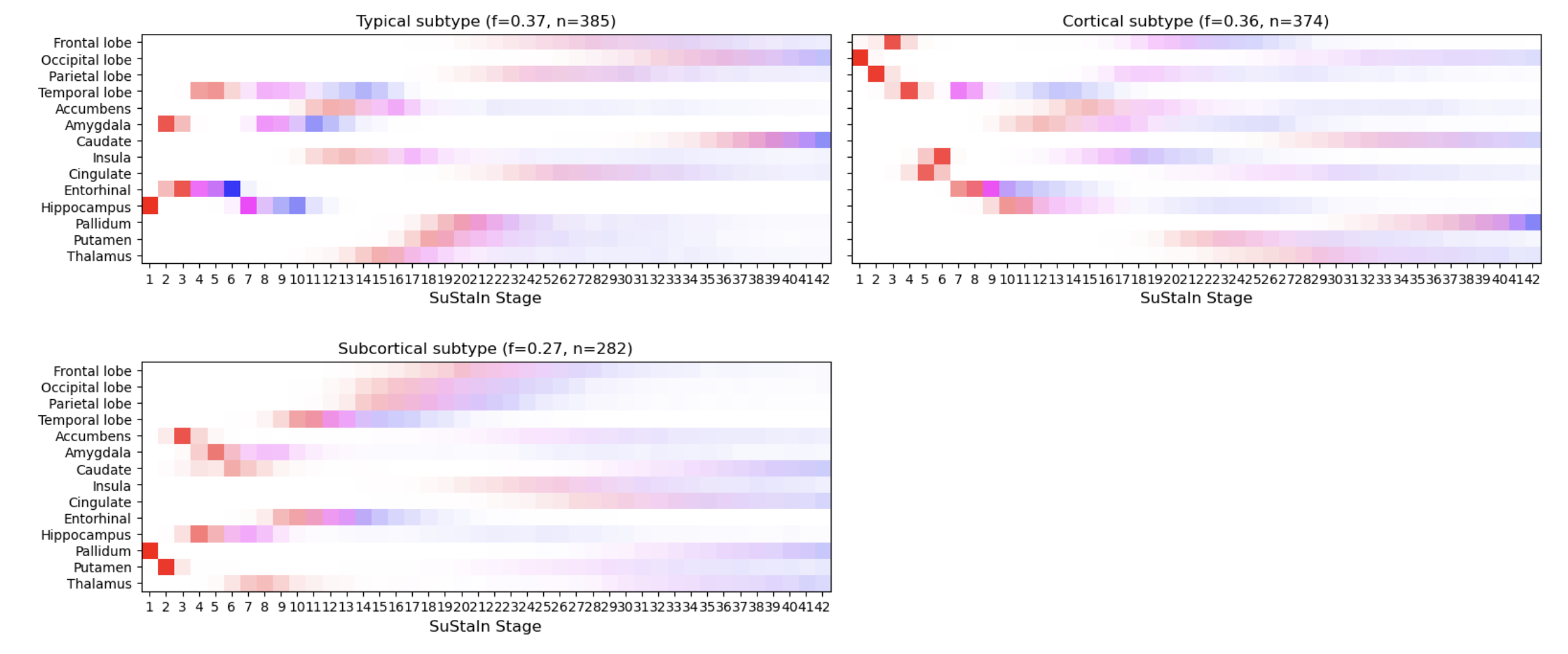}
\caption{\small \textbf{Results of experiment on ADNI using cortical thickness measuments for some ROIs.} The panels are PVDs \cite{EBM1} where the vertical axes shows the 14 regional brain volumes or thicknesses that reach atrophy z-scores along the x-axis of $z=1$ (red), $z=2$ (magenta) and $z=3$ (blue) in a probabilistic sequence (left-to-right). We used $CDR=0$ as control separator. We recognise the three original SuStaIn subtypes \cite{ZScoreSustain},  with slightly later atrophies in the Cortical subtype as well as less assigned individuals.   \\\textit{Abbreviations: PVD --- Positional Variance Diagram; CDR --- Clinical Demential Rating.}}
\label{fig:justCDRthick}
\end{figure}

Figure \ref{fig:justCDRvolumes} shows the stratification when using only volumes. Figure \ref{fig:justCDRthick} instead show the stratification when using cortical volumes and cortical thicknesses for the above mentioned ROIs.

As expected, we observe slightly earlier atrophy in the ROIs where cortical thickness was used instead of volume. However, the overall order of events remains unchanged, and we continue to identify the three original subtypes. The only notable difference is that when using volume data, fewer individuals are classified into the Cortical subtype, with many shifting to the Typical subtype. This may be due to the fact that the Cortical subtype is characterised by early atrophy in regions where thickness data were used. Since cortical thinning is one of the earliest detectable signs of cognitive decline \cite{pacheco2015greater}, individuals showing cortical atrophy based on thickness may be undergoing early thinning that hasn't yet affected volume. This raises the question of whether using thickness measurements could enable earlier detection of atrophy patterns and might be preferable to volume-based analysis. Future research could explore this possibility further.

\newpage

\subsection*{Left vs. right hemisphere experiments}

\begin{figure}[H]
\centering
\includegraphics[width=\textwidth]{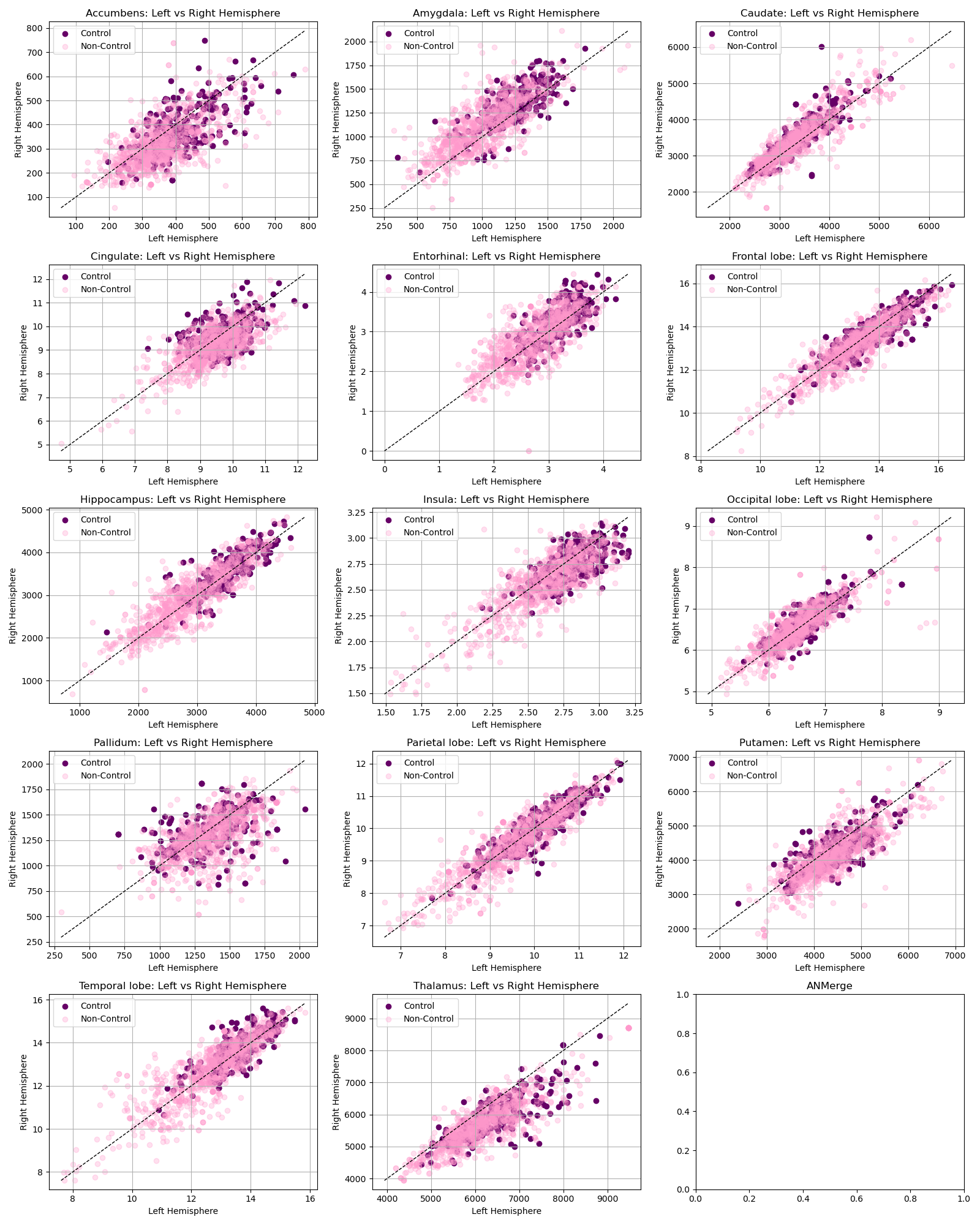}
\caption{\small \textbf{Scatter plots of left vs. right hemisphere ROIs measurements for the ANMerge cohort} The only region in which we observe a slight imbalance is the Thalamus. However, this region was not originally included in the SuStaIn study and is not characteristic of any subtype. }
\label{fig:hemisphere_comparison_diagnosis_anmerge}
\end{figure}

\newpage

\newpage

\newpage

\newpage

\subsection*{Additional visualisations}

\begin{enumerate}
\item ANMerge model NOT including control population

\begin{figure}[H]
\centering
\includegraphics[width=1\textwidth]{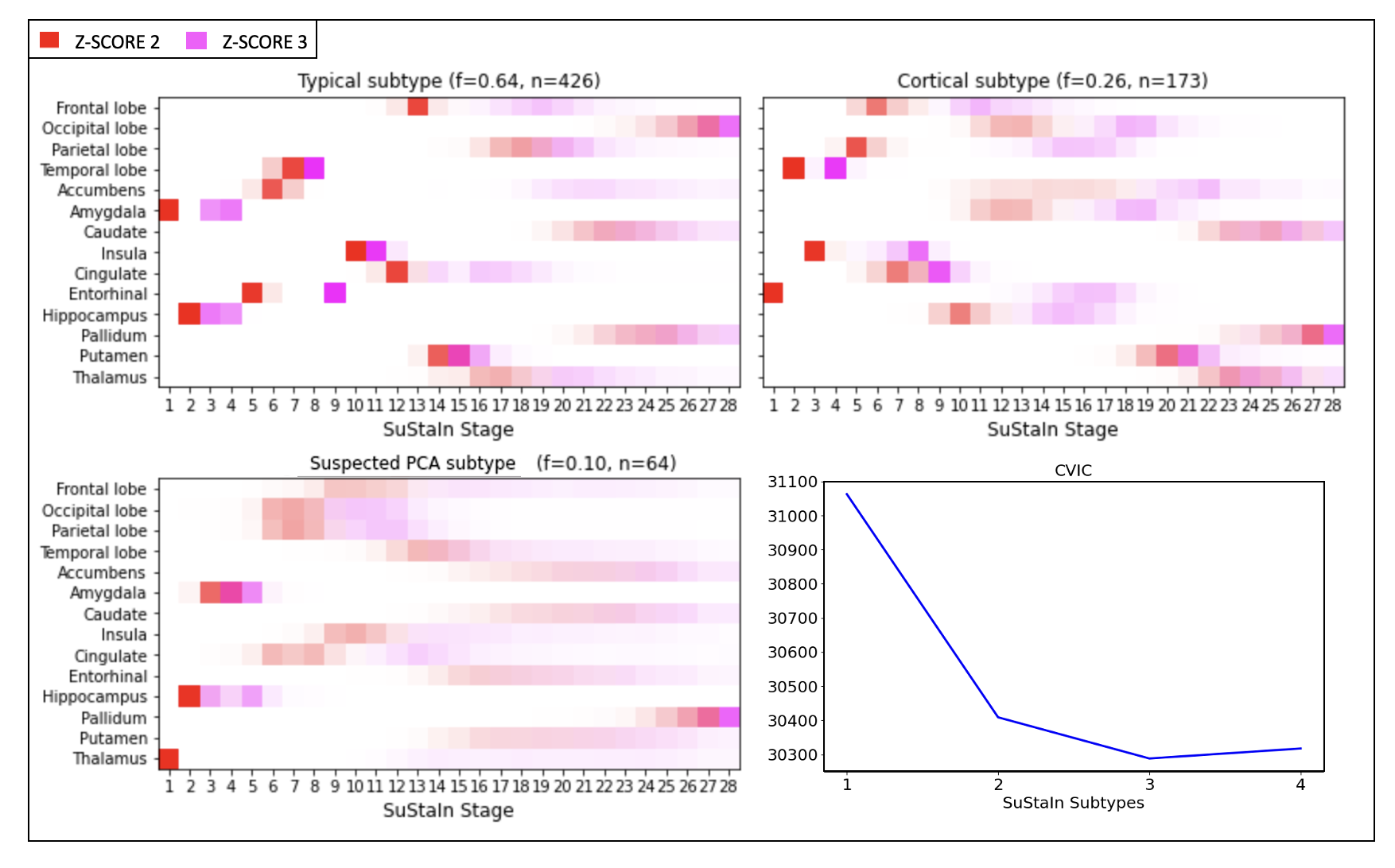}
\caption{\small \textbf{Results of experiment 2: 3-subtype atrophy model in 1.5T MRI, CN+MCI+AD participants, NOT including the control population, ANMerge dataset.} The lower right panel shows that optimum CVIC occurs for $N=3$ subtypes. The remaining panels are positional variance diagrams \cite{EBM1} (PVDs), where the vertical axes show the 14 input features (regional brain volumes) that reach atrophy z-scores along the x-axis of $z=2$ (red) and $z=3$ (magenta) in a probabilistic sequence (left-to-right). f and n indicate the average fraction and number of individuals assigned to each subtype across MCMC samples. The uncertainty of each event in the sequence can be inferred by the blurriness of the PVD. The PVDs shown in this figure are almost identical to the ones found by SuStaIn in the full cohort (Fig. \ref{fig:anmerge}). The Typical subtype also involved the Entorhinal cortex. The suspected PCA PVD is more blurry in this model, which may be due to the few individuals subtyped in this cluster. As for the Cortical subtype, the atrophy in the Cingulate and Insula dominates the sequence more strongly than the corresponding subtype with controls. \\\textit{Abbreviations: PCA --- posterior cortical atrophy; CVIC --- Cross Validation Information Criterion; CN --- Cognitively Normal; MCI --- Mild Cognitive Impairments; AD --- Alzheimer's Disease.}}
\label{fig:anmergeNC}
\end{figure}

\newpage

\item ADNI1.5T/ANMerge model NOT including control population

\begin{figure}[H]
\centering
\includegraphics[width=1\textwidth]{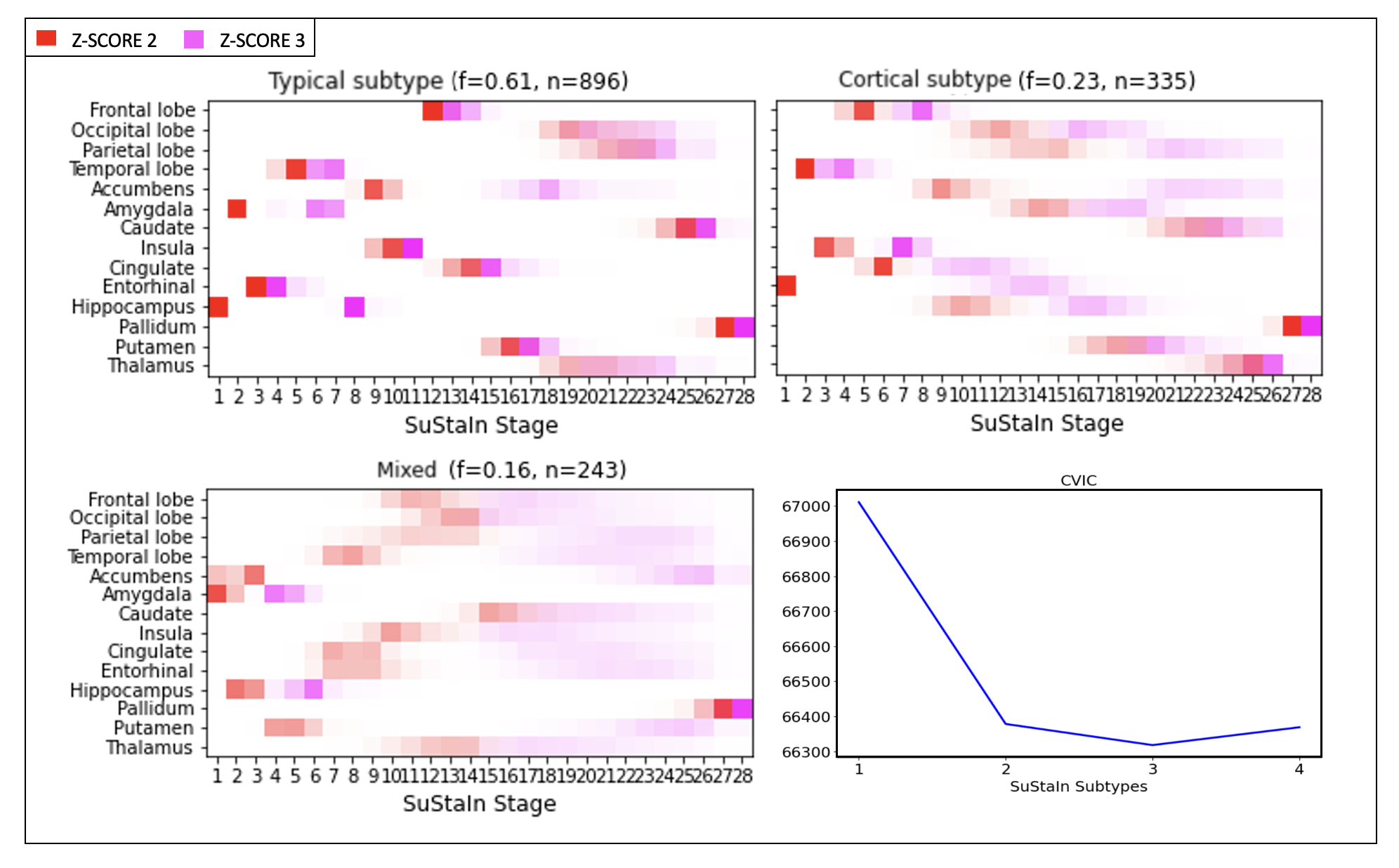} 
\caption{\small \textbf{Results of experiment 4: 3-subtype atrophy model in 1.5T MRI, CN+MCI+AD participants, NOT including the control population, ADNI and ANMerge dataset.} The lower right panel shows that optimum CVIC occurs for $N=3$ subtypes. The remaining panels are positional variance diagrams \cite{EBM1} (PVDs), where the vertical axes show the 14 input features (regional brain volumes) that reach atrophy z-scores along the x-axis of $z=2$ (red) and $z=3$ (magenta) in a probabilistic sequence (left-to-right). f and n indicate the average fraction and number of individuals assigned to each subtype across MCMC samples. The uncertainty of each event in the sequence can be inferred by the blurriness of the PVD.  The Typical and Cortical subtype strongly emerged in this model, both involving the Frontal Lobe.  \\\textit{Abbreviations: PCA --- posterior cortical atrophy; CVIC --- Cross Validation Information Criterion; CN --- Cognitively Normal; MCI --- Mild Cognitive Impairments; AD --- Alzheimer's Disease.}}
\label{fig:adnianmergeNC}
\end{figure}

\newpage

\item OASIS model including control population

\begin{figure}[H]
\centering
\includegraphics[width=1\textwidth]{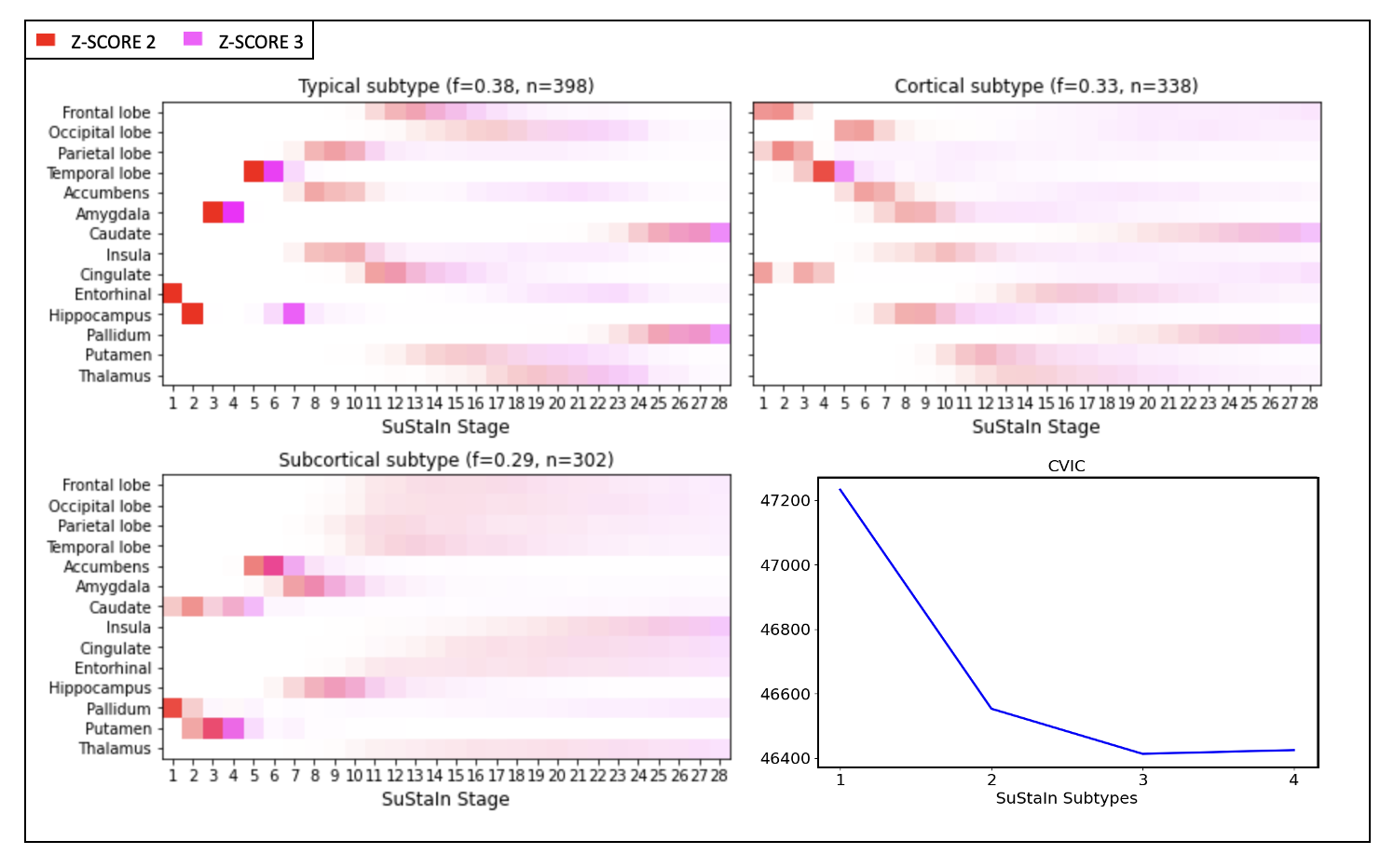}
\caption{\small \textbf{Results of experiment 5: 3-subtype atrophy models in 3.0T MRI, OASIS dataset.} The lower right panel shows that optimum CVIC occurs for $N=3$ subtypes. The remaining panels are positional variance diagrams \cite{EBM1} (PVDs), where the vertical axes show the 14 input features (regional brain volumes) that reach atrophy z-scores along the x-axis of $z=2$ (red) and $z=3$ (magenta) in a probabilistic sequence (left-to-right). f and n indicate the average fraction and number of individuals assigned to each subtype across MCMC samples. The uncertainty of each event in the sequence can be inferred by the blurriness of the PVD. All the three original patterns of atrophy were found in this model. \\\textit{Abbreviations: PCA --- posterior cortical atrophy; CVIC --- Cross Validation Information Criterion; CN --- Cognitively Normal; MCI --- Mild Cognitive Impairments; AD --- Alzheimer's Disease.}}
\label{fig:oasisC}
\end{figure}

\newpage

\item OASIS models NOT including control population

\begin{figure}[H]
\centering
\includegraphics[width=1\textwidth]{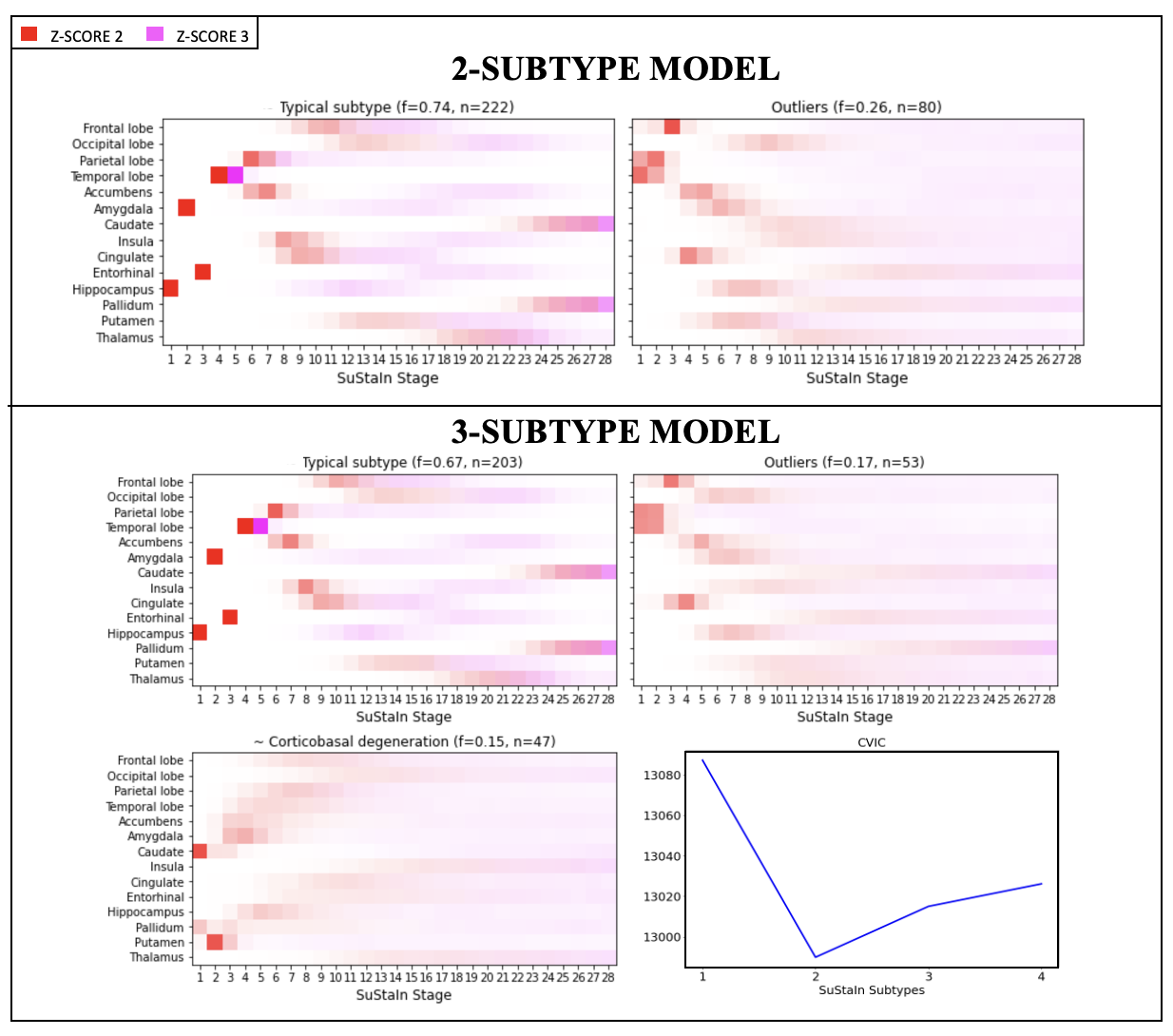}
\caption{\small \textbf{Results of experiment 6: 2-subtype and 3-subtype atrophy models in 3.0T MRI, NOT including the control population, OASIS dataset.} The lower right panel shows that optimum CVIC occurs for $N=2$ subtypes. The remaining panels are positional variance diagrams \cite{EBM1} (PVDs), where the vertical axes show the 14 input features (regional brain volumes) that reach atrophy z-scores along the x-axis of $z=2$ (red) and $z=3$ (magenta) in a probabilistic sequence (left-to-right). f and n indicate the average fraction and number of individuals assigned to each subtype across MCMC samples. The uncertainty of each event in the sequence can be inferred by the blurriness of the PVD.  We have included both the 2-subtype and 3-subtype models. What characterises these PVDs is the extremely high uncertainty associated with the event positioning. The Typical subtype emerged in both models, together with a spurious subtype with a few outliers showing several mild atrophies at early stages. It's worth reminding that the third subtype (suspected Corticobasal degeneration) was not cross validated as it did not improve the overall model likelihood. However, those 47 individuals show a very interesting pattern of atrophy in the Caudate and Putamen. These subjects could either exhibit an underlying disorder associated with some form of Corticobasal degeneration, or they could be outliers.  \\\textit{Abbreviations: PCA --- posterior cortical atrophy; CVIC --- Cross Validation Information Criterion; CN --- Cognitively Normal; MCI --- Mild Cognitive Impairments; AD --- Alzheimer's Disease.}}
\label{fig:oasisNC}
\end{figure}

\newpage

\item ADNI3T/OASIS models NOT including control population

\begin{figure}[H]
\centering
\includegraphics[width=0.9\textwidth]{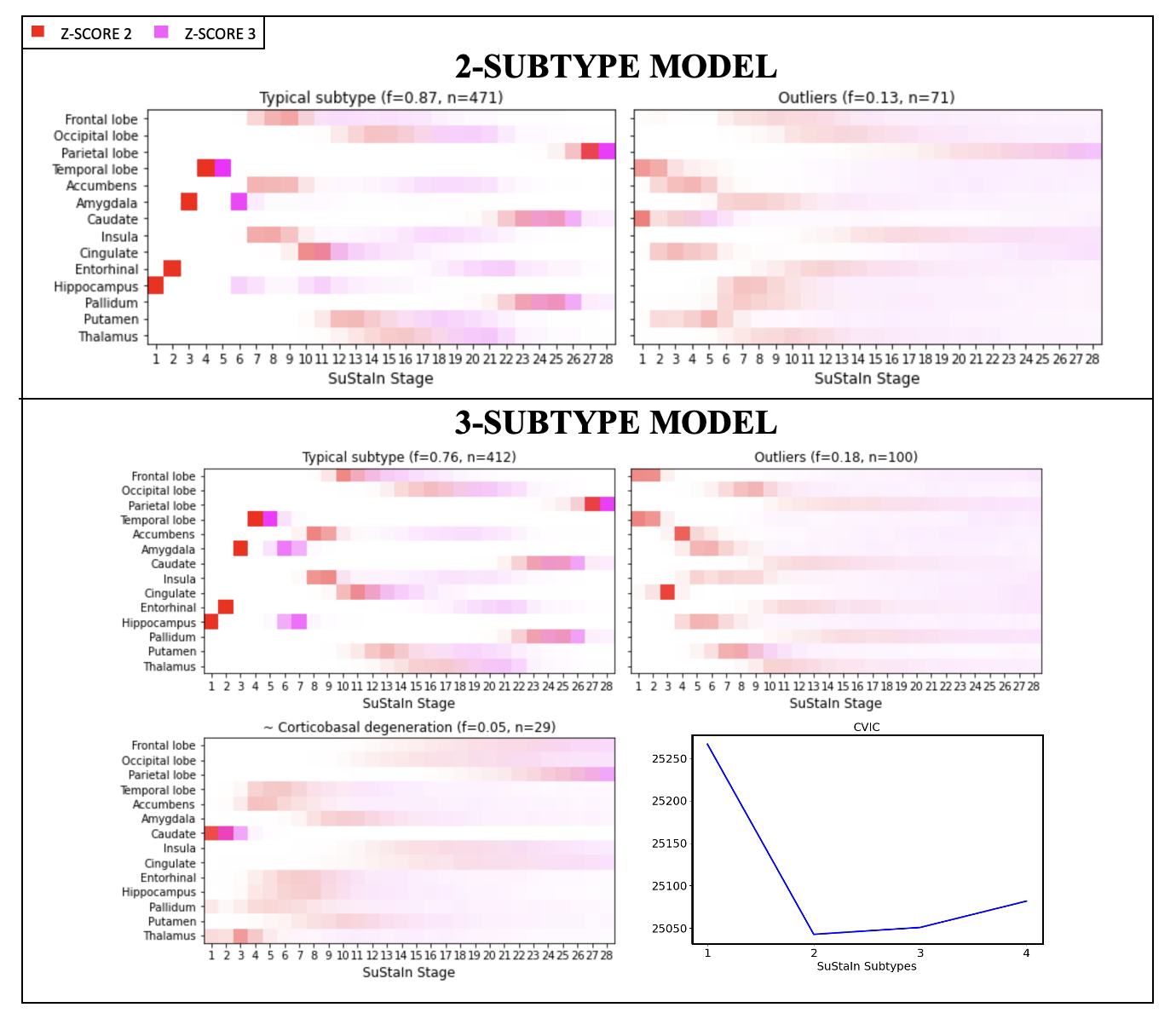}
\caption{\small \textbf{Results of experiment 8: 2-subtype and 3-subtype atrophy models in 3.0T MRI, NOT including the control population, ADNI and OASIS dataset.} The lower right panel shows that optimum CVIC occurs for $N=2$ subtypes. The remaining panels are positional variance diagrams \cite{EBM1} (PVDs), where the vertical axes show the 14 input features (regional brain volumes) that reach atrophy z-scores along the x-axis of $z=2$ (red) and $z=3$ (magenta) in a probabilistic sequence (left-to-right). f and n indicate the average fraction and number of individuals assigned to each subtype across MCMC samples. The uncertainty of each event in the sequence can be inferred by the blurriness of the PVD.  We have included both the 2-subtype and 3-subtype models. The Typical subtype clearly emerged in both models, together with a spurious subtype with a few outliers  showing several mild atrophies at early stages. It's worth reminding that the third subtype (suspected Corticobasal degeneration) was not cross validated as it did not improve the overall model likelihood. However, this third subtype not only separated individuals with very strong and early Caudate atrophy, it also improved the disease progression sequence of the second cluster, which started to show a cortical pattern. Again, the severe shrinking of the Caudate could be associated with some form of Corticobasal degeneration. It is promising that SuStaIn was able to separate very few subjects with an extremely specific atrophy. \\\textit{Abbreviations: PCA --- posterior cortical atrophy; CVIC --- Cross Validation Information Criterion; CN --- Cognitively Normal; MCI --- Mild Cognitive Impairments; AD --- Alzheimer's Disease.}}
\label{fig:adnioasisNC2}
\end{figure}

\end{enumerate}

\newpage

\begin{table}[]
\centering
\setlength{\tabcolsep}{15pt}
\begin{tabular}{p{0.1\textwidth}p{0.3\textwidth}p{0.4\textwidth}}
\toprule
Domain  & ADAS-Cog &  MMSE   \\
\midrule
\midrule
Vision & \begin{itemize}
    \item \textbf{Task 2} patients were shown objects and/or fingers and had to name them
    \item \textbf{Task 4} patients asked to copy geometric forms from drawings/pictures
\end{itemize} &  \begin{itemize}
    \item What is this? (patients shown object)
    \item Please copy the drawing
\end{itemize}  \\
\midrule
Time & \begin{itemize}
    \item \textbf{Task 6} patients asked time of the day, date, month, season, year etc.
\end{itemize} &  \begin{itemize}
    \item What is the day of the week?
    \item What is the date
    \item What is the month?
    \item What is the season?
    \item What is the year?
\end{itemize}\\
\midrule
Space & \begin{itemize}
    \item \textbf{Task 6} patients asked to identify place, route, locate home
\end{itemize} &  \begin{itemize}
    \item What is the name of this place/hospital?
    \item Could you name two nearby streets?
    \item What is the name of this town?
    \item What is the name of this district?
    \item What floor of the building are we on?
\end{itemize} \\
\midrule
Spelling &  &  Spell WORLD backwards \\
\midrule
Maths &  &  Can you take 7 away from 100, then add 6 ... \\
\bottomrule
\bottomrule
\end{tabular}
\captionsetup{justification=centering}
\caption{\small \label{ANMergecogass} \textbf{ANMerge cognitive assessments.}  This table highlights the main cognitive domains examined for the ANMerge cohort and the particular single questions/tasks chosen.} 
\end{table}

\newpage

\begin{figure}[H]
\centering
\includegraphics[width=1\textwidth]{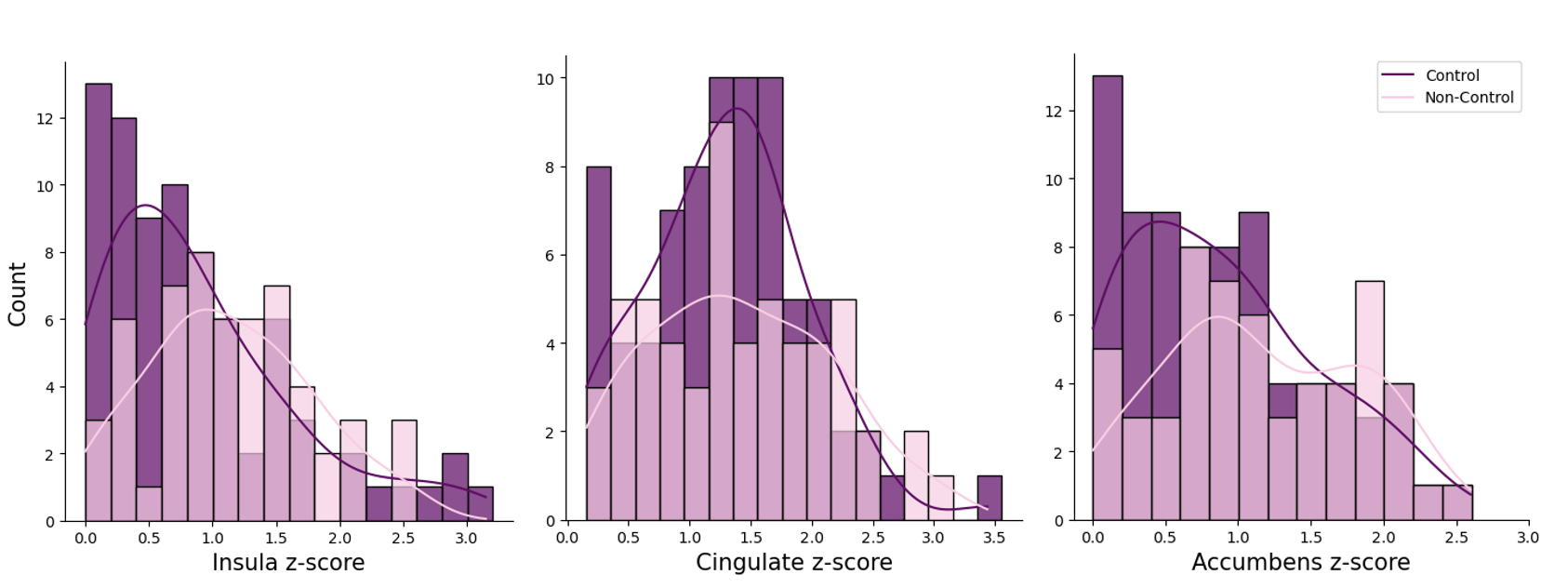}
\caption{\small \label{fig:z-score dist cortical}\textbf{Z-score distribution of OASIS individuals assigned to cortical subtype.} This figure shows how the z-scores are distributed in the individuals assigned to the cortical subtype in OASIS. We do observe a difference between control and non-control subjects but they both reach very high z-scores.  }
\label{fig:z-score dist cortical}
\end{figure}

\begin{figure}[H]
\centering
\includegraphics[width=1\textwidth]{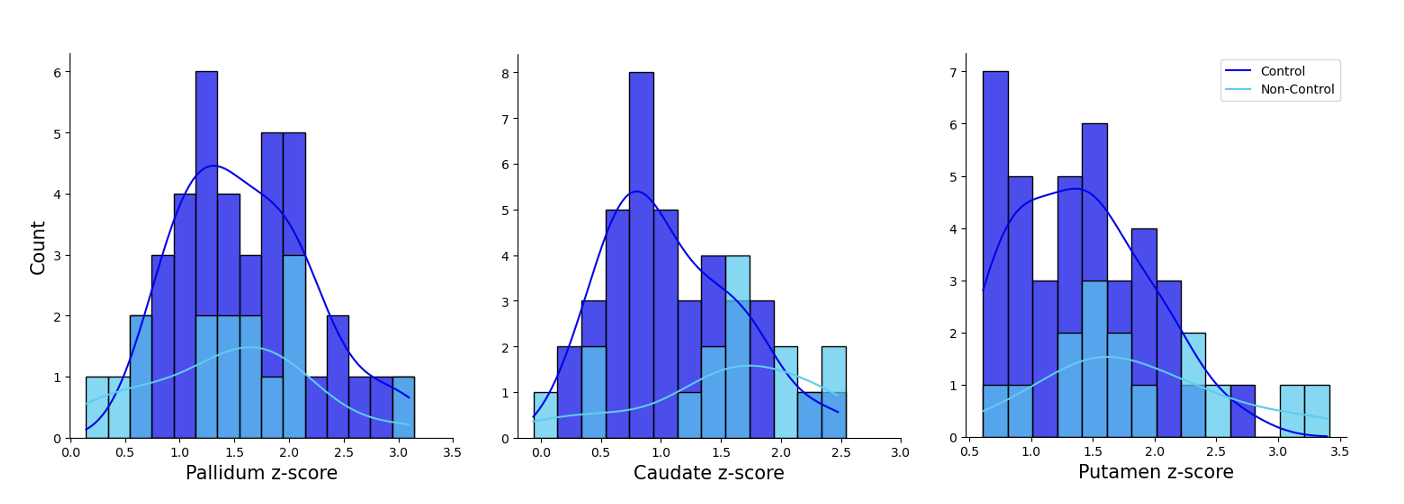}
\caption{\small \label{fig:z-score dist}\textbf{Z-score distribution of OASIS individuals assigned to Subcortical subtype.} Distribution of z-scores in the individuals assigned to the Subcortical subtype in OASIS. There is no substantial difference between control and non-control subjects as they both reach very high z-scores.}
\label{fig:z-score dist}
\end{figure}

\begin{figure}[H]
\centering
\includegraphics[width=1\textwidth]{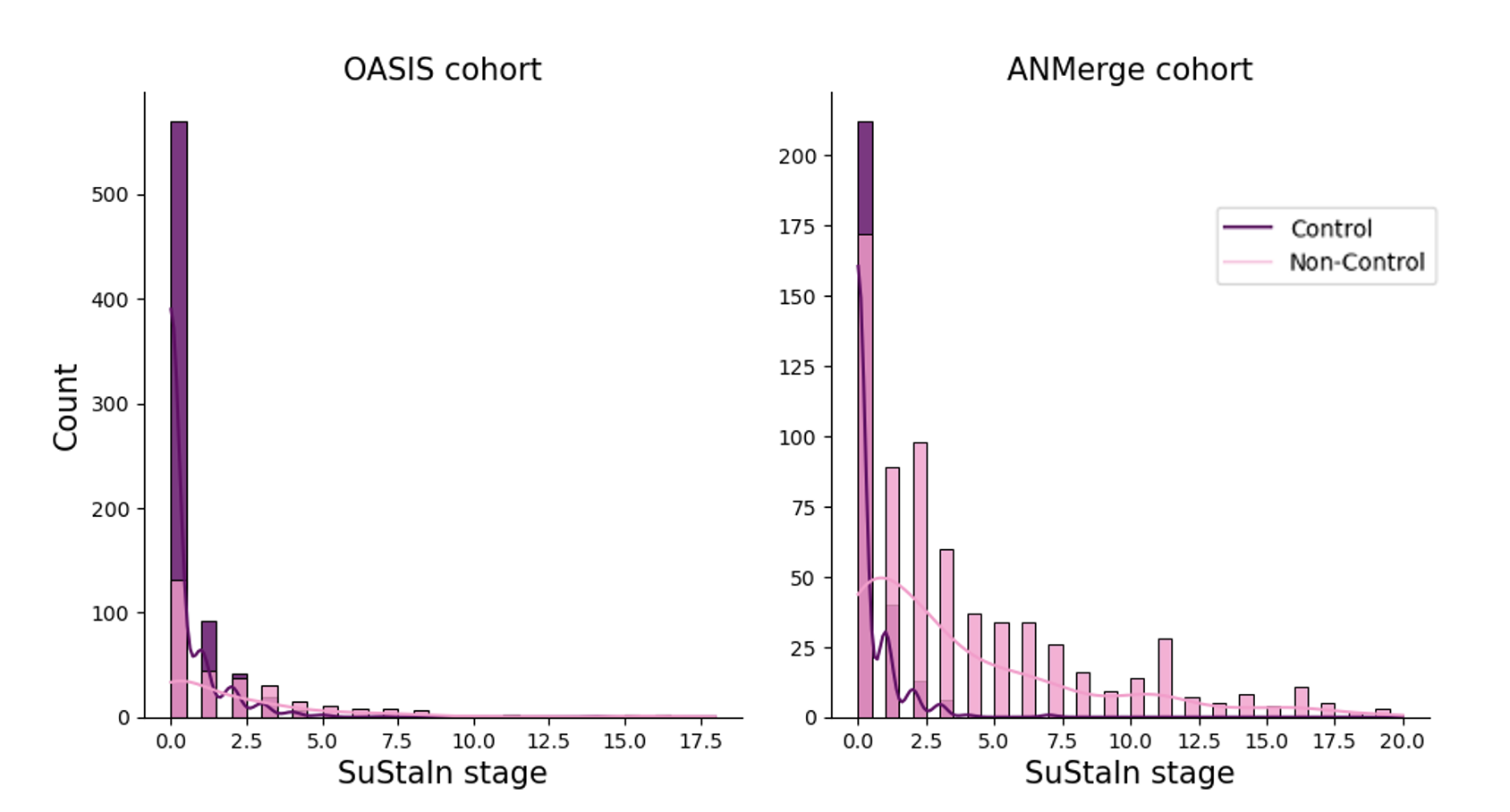}
\caption{\small \label{fig:stages oasis anmerge}\textbf{Histograms of SuStaIn disease stage assignment across OASIS and ANMerge individuals. }  It is evident in the OASIS histogram that the majority of the individuals have a CDR score of zero and are staged at zero. However, looking at the curves, while there is a significant difference between the distribution of controls and non-controls in ANMerge, with the latter reaching higher stages, for OASIS individuals the staging is very similar across these two groups.}
\label{fig:stages oasis anmerge}
\end{figure}

\begin{figure}[H]
\centering
\includegraphics[width=1\textwidth]{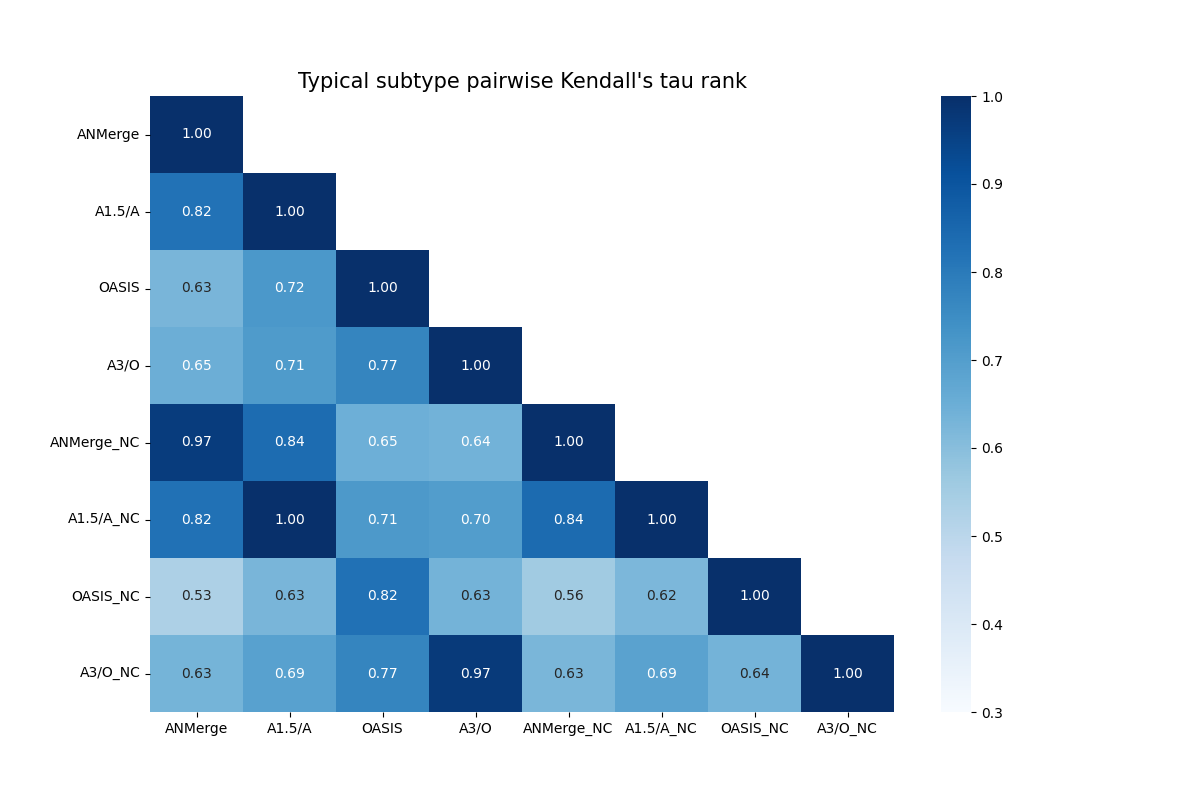}
\caption{\small \label{fig:typical tau}\textbf{Pairwise Kendall's tau rank between typical subtypes across different models. }Given the typical subtype emerged in all 8 experiments, it is possible to compare the disease sequences across all of them. We observe a generally high sequence concordance.}
\label{fig:typical tau}
\end{figure}

\begin{figure}[H]
\centering
\includegraphics[width=1\textwidth]{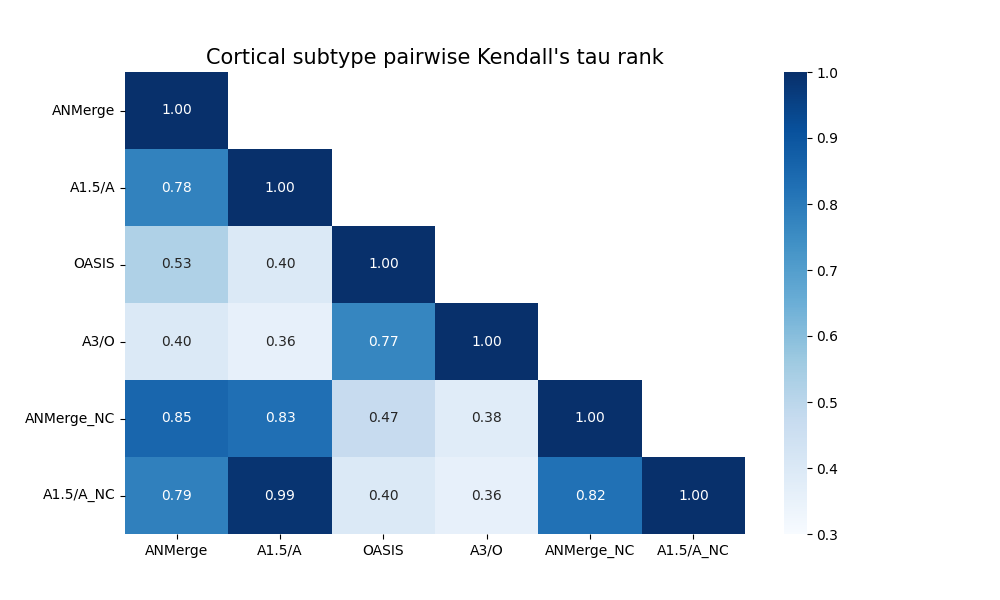}
\caption{\small \label{fig:cortical tau}\textbf{Pairwise Kendall's tau rank between cortical subtypes across different models } A few experiments are missing because the cortical subtype did not emerge in those models. The sequence concordance is generally very high except between cohorts with no overlapping data and different magnetic field strengths, though the ranks still indicate good concordance. }
\label{fig:cortical tau}
\end{figure}

\begin{figure}[H]
\centering
\includegraphics[width=1\textwidth]{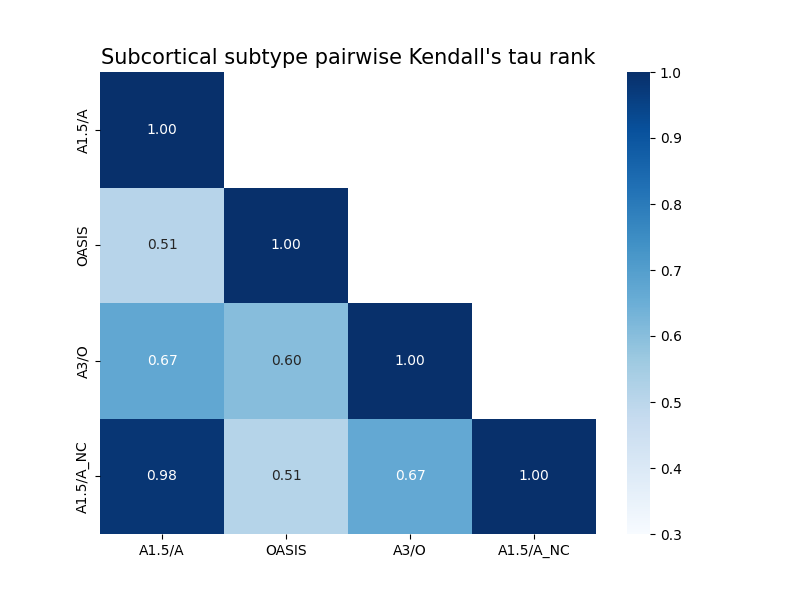}
\caption{\small \label{fig:subcortical tau}\textbf{Pairwise Kendall's tau rank between subcortical subtypes across different models } Half of the experiments are missing because the subcortical subtype did not emerge in those models. The sequence concordance is generally very high. }
\label{fig:subcortical tau}
\end{figure}

\end{document}